\documentclass[pra,twocolumn,english,showpacs,floatfix,superscriptaddress]{revtex4-1}
\usepackage{graphicx}
\usepackage{amssymb}
\usepackage{color}
\usepackage{amsmath}

\begin{document}
\title{Accessing finite momentum excitations of the one-dimensional Bose-Hubbard model using superlattice modulation spectroscopy}

\author{Karla Loida}
\affiliation{HISKP, University of Bonn, Nussallee 14-16, 53115 Bonn, Germany}
\author{Jean-S\'ebastien Bernier}
\affiliation{HISKP, University of Bonn, Nussallee 14-16, 53115 Bonn, Germany}
\author{Roberta Citro}
\affiliation{Dipartimento di Fisica “E.R. Caianiello“, Universit\`a degli Studi di Salerno, Via Giovanni Paolo II 132, I-84084 Fisciano (Sa), Italy}
\author{Edmond Orignac}
\affiliation{Universit\'e de Lyon, \'Ecole Normale Sup\'erieure de Lyon, Universit\'e Claude Bernard, CNRS, Laboratoire de Physique, F-69342 Lyon, France}
\author{Corinna Kollath}
\affiliation{HISKP, University of Bonn, Nussallee 14-16, 53115 Bonn, Germany}

\begin{abstract}
  We investigate the response to superlattice modulation of a bosonic quantum gas confined to arrays of
  tubes emulating the one-dimensional Bose-Hubbard model. We demonstrate, using both time-dependent density matrix
  renormalization group and linear response theory, that such a superlattice modulation gives access to the excitation spectrum
  of the Bose-Hubbard model at finite momenta. Deep in the Mott-insulator, the response is characterized by a
  narrow energy absorption peak at a frequency approximately corresponding to the onsite interaction strength between bosons.
  This spectroscopic technique thus allows for an accurate measurement of the effective value of the interaction strength.
  On the superfluid side, we show that the response depends on the lattice filling. The system
  can either respond at infinitely small values of the modulation frequency or only above a frequency threshold.
  We discuss our numerical findings in light of analytical results obtained for the Lieb-Liniger model. In particular,
  for this continuum model, bosonization predicts power-law onsets for both responses.
\end{abstract}


\date{\today}
\maketitle

The one-dimensional Bose-Hubbard model, one of the most celebrated models of many-body quantum physics, 
describes the intriguing interplay of quantum kinetic processes and local interaction.
Although conceptually simple, this model is not exactly solvable even in one dimension, but, thankfully,
due to years of hard work its ground state phase diagram is now
well understood~(\cite{KuehnerMonien1998,KuehnerMonien2000,ZakrzewskiDelande2008,BarmettlerKollath2012,CazalillaRigol2011,LaeuchliKollath2008}
and references therein). 
For commensurate filling, an
interaction-induced Mott insulator and a superfluid state are known to be separated
at a critical value of the interaction strength by a quantum phase transition of the
Kosterlitz-Thouless type. While for incommensurate filling, the system remains superfluid for
arbitrary interaction strength. Since the first realization of the Bose-Hubbard model
using ultracold atoms in optical lattices more than a decade ago~\cite{JakschZoller1998,GreinerBloch2002},
various experimental verifications of the properties of the one-dimensional model have been carried out~
\cite{BlochZwerger2008,ParedesBloch2004,StoeferleEsslinger2004,KoehlEsslinger2005_2,ClementInguscio2009,HallerNaegerl2010}.
Despite these advances, fully understanding the excitation spectrum of the Bose-Hubbard model still
requires more work. In this light, the development of powerful techniques to probe the excitations of
cold atom systems is extremely promising. Particularly useful are
spectroscopic methods such as radio frequency, Raman, Bragg or lattice modulation
spectroscopy which give access to single-particle, density or kinetic energy
spectral functions~\cite{BlochZwerger2008,HallerNaegerl2010,Toermae2015,BissbortHofstetter2011,FabbriCaux2015,StoeferleEsslinger2004,MeinertNaegerl2015,KoehlEsslinger2005_2}.

The latter method, lattice modulation spectroscopy, measures the response of a system to a time-dependent
modulation of the lattice amplitude. In bosonic gases, the energy added to the system due to the modulation
is extracted from the broadening of the central momentum peak in a time-of-flight measurement.
Lattice modulation spectroscopy was first introduced to characterize the excitations
across the phase transition between the superfluid and Mott-insulating states and
has been applied to different geometries including one-dimensional lattices~\cite{StoeferleEsslinger2004,HallerNaegerl2010}.
A sizable corpus of theoretical studies have shown that for Bose-Hubbard systems this measurement technique is an adequate
probe of the excitations at zero quasi-momentum transfer~\cite{MenottiStringari2003,BatrouniDenteneer2005,ReischlUhrig2005,
  PupilloBatrouni2006,KraemerDalfovo2005,IucciGiamarchi2006,KollathSchollwoeck2006, SensarmaDasSarma2011}. 
Moreover, lattice modulation spectroscopy was employed to study strongly interacting bosons loaded
in disordered lattices~\cite{FallaniInguscio2007,OrsoGiamarchi2009}, and to reveal signatures of the Higgs mode in the
two-dimensional superfluid system near the transition to the Mott phase~\cite{HuberBlatter2007, EndresBloch2012}.

\begin{figure}
\includegraphics[width=.99\columnwidth,clip=true]{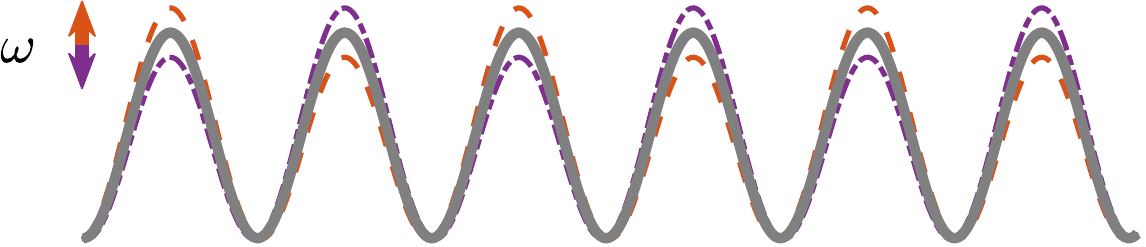}
\caption{Sketch of the superlattice modulation spectroscopy. The amplitude of the equilibrium optical
  lattice $V_0(x)=V_0\sin^2(k_Lx)$ (gray solid line) is time-periodically modulated in a dimerized fashion,
  i.e. the perturbing potential is given by  $\delta V(x,t)\approx A\sin(\omega t)\sin(k_Lx)$ with small amplitude $A$.
  The lattice amplitude is modulated between the two configurations indicated by (orange) dashed and (purple) dash-dotted lines,
  illustrating that while one potential barrier is increased the neighboring one is decreased and vice versa.}
\label{fig:modulation}
\end{figure}

However, most of the previous lattice modulation setups only considered excitations
at low momenta as standard lattice modulation spectroscopy conserves quasi-momentum. 
Here we propose instead to use {\it superlattice} modulation spectroscopy to probe the excitation spectrum
of the Bose-Hubbard model at finite momenta. Superlattice modulation spectroscopy has recently been proposed
in fermionic systems as a technique to measure the temperature of a non-interacting system~\cite{LoidaKollath2015}
and to detect signatures of the exotic bond order wave phase in the ionic Hubbard model~\cite{LoidaKollath2017}.
In contrast with standard lattice amplitude modulation,
this approach modulates the lattice amplitude in a dimerized fashion such that
a finite momentum $\pi/a$ is transferred to the atoms (where $a$ is the lattice spacing).
To do so, one should choose the superlattice configuration such that 
the bottom offsets stay approximately constant corresponding to a dimerized modulation of the hopping amplitude.
Experimentally, the parameters of the laser beams forming the optical superlattice configuration can be fine tuned such that the
equilibrium lattice is approximated by the simple form $V_0(x)=V_0\sin^2(k_Lx)$. Additionally, a time-periodic and site alternating
modulation of the lattice height $\delta V(x,t) \approx A\sin(\omega t)\sin(k_Lx)$ (for small amplitude $A$) can
be engineered by periodically tuning in time the phase between the laser waves generating the optical superlattice.
Here $k_L$ is the magnitude of the wave vector of the lattice light.

We study here the response of the one-dimensional Bose-Hubbard model to superlattice modulation using
time-dependent density matrix renormalization group method (t-DMRG)~\cite{Schollwoeck2011} and linear response theory, the latter
approach being combined with perturbation theory for strong interaction strengths and bosonization for weak interaction strengths.
We demonstrate that the absorbed energy as a response to superlattice modulation provides precise information
on the excitation spectrum at finite momenta for both the superfluid and Mott-insulating phases.
In the Mott insulator, we find a narrow and distinct absorption peak at a modulation frequency $\hbar \omega \sim U$ enabling
for a precise calibration of the interaction strength $U$.
While on the superfluid side, we show that depending on the lattice filling the system
can either respond at infinitely small values of the modulation frequency or only above a frequency threshold.
This behavior highlights the correspondence between
the low energy spectral features of the weakly interacting Bose-Hubbard superfluid
and those of the Lieb-Liniger model~\cite{Lieb1963,LiebLiniger1963}.

The rest of this article is organized as follows. In Section~\ref{sec:setup}, we introduce
the theoretical framework. We define the equilibrium system and the superlattice amplitude
modulation. We then introduce the quasi-exact time-evolution used to compute the observable of interest,
the absorbed energy, and we show how this quantity relates to the averaged energy absorption rate
within linear response theory. In Section~\ref{sec:MI}, we investigate the response of the Mott insulator to
superlattice modulation. We first introduce in Section~\ref{sec:MI_PT} an analytical approach based on linear
response and perturbation theory valid for large interaction strengths, and then in Section~\ref{sec:MI_results}
compare our analytical predictions to the numerical results obtained using t-DMRG.
In Section~\ref{sec:SF}, we investigate the response of the superfluid to superlattice modulation spectroscopy.
We first present in Section~\ref{sec:SF_LL} the excitation spectrum expected within the Lieb-Liniger and
Luttinger liquid theories before discussing how to probe the continuous and gapped parts of the spectrum.
In the subsequent Section~\ref{sec:SF_results}, we present the corresponding numerical results obtained using
t-DMRG. Finally, we conclude in Section~\ref{sec:con}.

\section{Setup and theoretical model}\label{sec:setup}
We consider ultracold bosonic atoms confined to one-dimensional tubes which can for example be realized using a strong
two-dimensional optical lattice perpendicular to the tube direction. Along the one-dimensional tube direction an additional
weaker lattice is applied creating a periodic potential for the bosonic atoms. For sufficiently deep lattices, each tube can
be described by the one-dimensional Bose-Hubbard model,
\begin{align}\label{eqn:H0}
H_0&=H_{\text{kin}}+H_U \nonumber \\
&=-J\sum _{j=1}^{L-1} (a_j^{\dag}a_{j+1}^{\phantom \dag}+\text{h.c.}) +\frac{U}{2}\sum _{j=1}^{L}n_j(n_j-1),
\end{align}
where $a_j$ and $a_j^{\dag}$ represent the bosonic annihilation and creation operators at site $j$ and $n_j=a^\dagger_j a_j$
is the local number operator, $L$ is the even number of lattice sites. The kinetic part of the Hamiltonian $H_{\text{kin}}$
has tunneling amplitude $J$ and the effective onsite interaction strength $U/J$ can be tuned over several orders of magnitude
by tuning the lattice height.

\begin{figure}
\includegraphics[width=.99\columnwidth,clip=true]{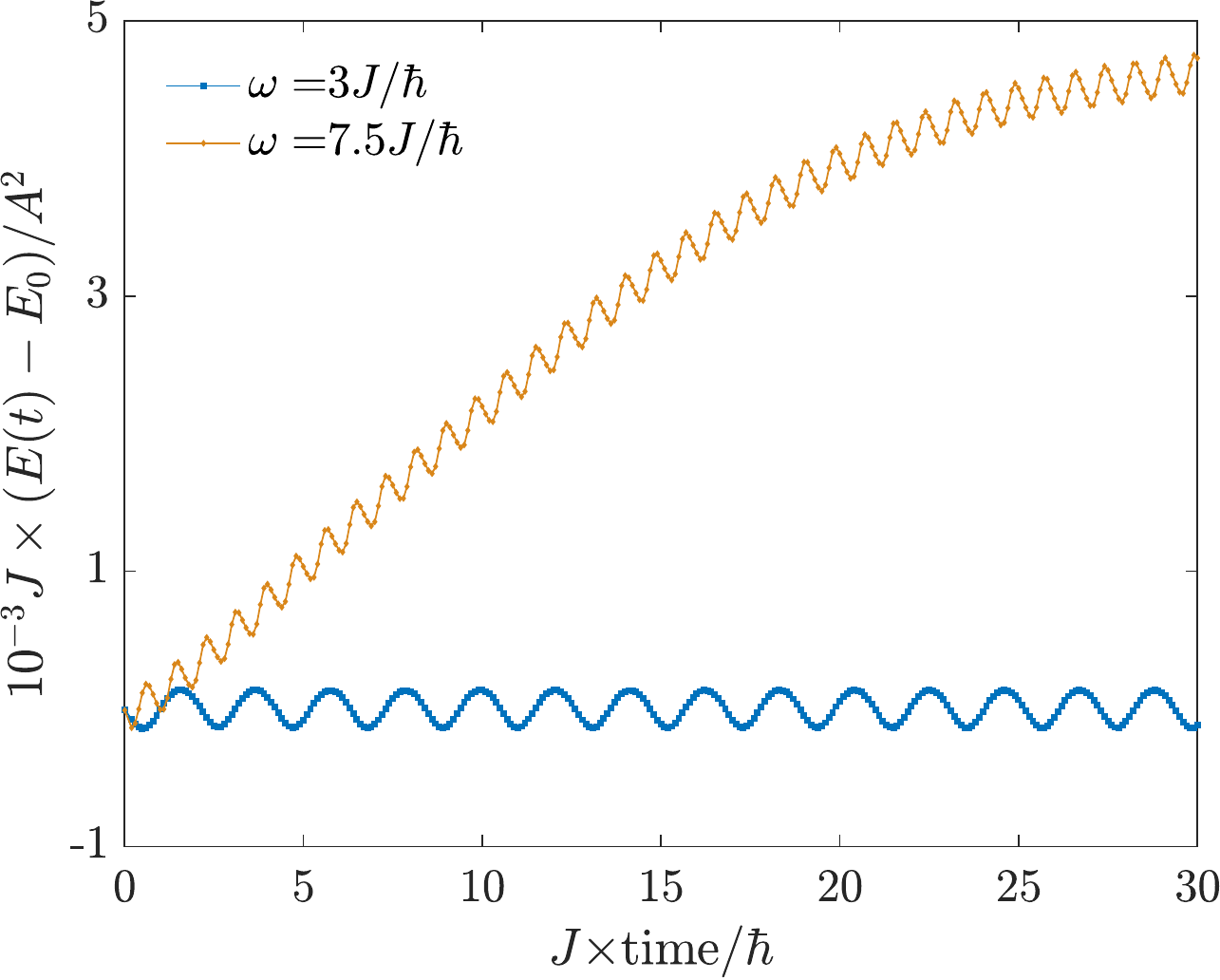}
\caption{Time evolution of the absorbed energy $E(t)-E_0$ at $U=4J$ in a system of $L=64$ sites and
  filling ${\bar n}=1$ per site for a modulation amplitude $A=0.01J$ and two different modulation frequencies.
  If this frequency is chosen within the resonant region, $\hbar\omega=7.5J$ (orange), energy is absorbed, whereas when
  the modulation is off-resonant, $\hbar\omega=3J$ (blue), very little energy absorption takes place.}
\label{fig:Et}
\end{figure}

In order to create excitations with finite momentum, we apply an amplitude modulation in a superlattice geometry (see Fig.~\ref{fig:modulation}).
The modulation is chosen in such a way that the bottoms of the potential wells are fixed while their heights are
modulated in a dimerized fashion. This means that while the lattice height on one bond increases, it decreases on the
two neighboring bonds. For small modulations, this setup is described by a dimerized modulation of the hopping parameter, i.e.~the
perturbation can be described by $H_\text{\text{pert}}=A\sin(\omega t)\hat O$ where $A \ll J$ is a small
amplitude and $\omega$ is the frequency of the modulation and the perturbation operator
\begin{eqnarray}
  \hat O &=& \sum_{j=1}^{L-1}(-1)^j(a_j^{\dag}a_{j+1}^{\phantom \dag}+\text{h.c.})
  = 2i\sum_{k}\sin(k a)a_{{k}+\frac{\pi}{a}}^{\dag}a_{k}^{\phantom \dag}. \nonumber \\
  \label{eqn:O}
\end{eqnarray}
Here we used the Fourier transform of the bosonic creation operator $a_j^{\dag}=(1/\sqrt{L})\sum_k\exp(iajk)a_k^{\dag}$
and $k = 2\pi r/(L a)$ with $r = 0, ..., L-1$. Compared to normal lattice modulation and rf-spectroscopy
(which are momentum conserving) this operator transfers a finite momentum to the system as shown
in Eq.~\eqref{eqn:O}. 
In order to quantify the amount of excitations created, we monitor the time-evolution of the absorbed energy.
To do this we simulate numerically the time-evolution of the initial ground state of $H_0$ under
the Hamiltonian $H(t)=H_0+H_{\text{pert}}$. 
Typical evolutions of the absorbed energy are illustrated in Fig.~\ref{fig:Et}.
The results are obtained simulating the full time-dependent problem using
time-dependent density matrix renormalization group (t-DMRG) described
in Refs.~\cite{DaleyVidal2004,WhiteFeiguin2004}. If the system is perturbed at a frequency far from any
resonant excitation, the energy remains approximately constant with slight changes. However, 
if excitations can be created resonantly,
the energy absorption displays a steep linear rise followed by a saturation.
The linear rise can often be understood within linear response theory and, when
suitable, we will compare our simulations to analytical results obtained within this framework. 
To carry out the time-evolution using t-DMRG, we keep a matrix dimension of
$D=128$ and the local number of bosons is restricted to $\sigma=3$ for $U\geq 15J$ and $\sigma=7$
for $U\leq 10J$. We conduct an error analysis by increasing the matrix dimension to $196$ states and the
local number of bosons to $\sigma+2$. In the Trotter-Suzuki time evolution we set $J\Delta t=0.01\hbar$ (except for $L=96$ where
we set  $J\Delta t=0.005\hbar$) and we use $J\Delta t=0.005\hbar$ ($J\Delta t=0.001\hbar$) to perform
the error analysis. In the linear regime, fitting the time-dependent absorbed energy, we
extract the energy absorption rate. The error bars provided in the figures show the maximal uncertainty
due to the matrix dimension, the local boson number, the time-step and variations of the fit range.

Within linear response, the energy absorption rate at zero temperature (corresponding to the slope of
the linear rise of the energy) is
\begin{align}\label{eqn:dEdt}
  \frac{\overline{dE(t)}}{dt}(\omega) &=\frac{\pi}{2}\omega \vert A \vert^2 \sum_{\alpha}
  \vert \langle \alpha \vert \hat O\vert \text{GS}\rangle \vert^2 \delta \left (\hbar\omega +E_0 -E_\alpha\right ).
\end{align}
Here $E_\alpha$ are the eigenenergies of the unperturbed Hamiltonian $H_0$, $|\alpha \rangle$ the corresponding eigenstates,
and $\omega$ the modulation frequency. 
The $\delta$-function in this expression ensures that excitations are created resonantly: the excitation
energy provided by the modulation, $\hbar \omega$, needs to equal the difference between the ground state energy, $E_0$, and
one of the excited states $E_\alpha$. The amplitude of the created excitations is additionally set by the matrix
element of the perturbation operator $\hat O$ between the ground state of $H_0$, $|\text{GS}\rangle$,
and the excited state $|\alpha\rangle$. The difficulty of the application of this formula in a many-body context typically lies
in determining the eigenstates and their respective eigenenergies. 

\section{Response on the Mott-insulating side of the phase transition}\label{sec:MI}
In this section we discuss the response of a one-dimensional Mott-insulating state to the
superlattice modulation spectroscopy. We compare our numerical results to a perturbative approach in $J/U$ and
point out how our modulation scheme differs from normal lattice spectroscopy.

\subsection{Perturbation theory}\label{sec:MI_PT}
In the strong coupling limit of the Mott-insulating phase, we employ a perturbative approach considering the
first non-vanishing order in $J/U$ to evaluate the energy absorption rate within linear response using Eq.~\eqref{eqn:dEdt}.
We consider $H_U$ as the unperturbed Hamiltonian and $H_{\text{kin}}$ as the small perturbation as was performed
in Ref.~\cite{IucciGiamarchi2006} for the normal lattice amplitude modulation. We sketch here
the derivation for the superlattice modulation. 
\paragraph{Zeroth order}
At commensurate filling $\bar n$ per site, the ground state of the unperturbed Hamiltonian $H_U$ is given by
the atomic Mott-insulator $\vert 0 \rangle = \vert \bar n, \bar n, ..., \bar n\rangle$. For notational convenience,
we shift the energy scale such that the groundstate energy vanishes [$H_U=(U/2)\sum_j(n_j-\bar n)^2$], i.e.~$E_0 =0$,
and we consider a system with periodic boundary conditions. The excited states of $H_U$ lowest in energy
are created by a particle-hole excitation, i.e.~adding
one particle at a chosen site $m\in [1,...,L]$ and removing a particle from a different site $\tilde{m}$.
Here $\tilde{m}= m+d $ and $d\in [1,...,L-1]$ is the distance to the right from the site with occupation $\bar n+1$
to the site with occupation $\bar n-1$. This excited state can be written as
\begin{eqnarray}
  \vert m,d \rangle =\frac{1}{\sqrt{\bar n(\bar n+1)}}~a_{\tilde {m}}^{\phantom \dag}a_m^{\dag}\vert 0 \rangle \nonumber
\end{eqnarray}
and has eigenenergy $U$. Higher excited states have eigenenergies that are multiples of $U$. Due to
the high degeneracy of the excited states, one needs to employ degenerate perturbation theory~\cite{Sakurai1994}
in order to take into account the perturbation by the kinetic term $H_{\text{kin}}$. 
\paragraph{First order}
Up to first order, the ground state energy remains zero, while the correction to the ground
state wave function is
\begin{eqnarray}
  \vert \Psi_0^1\rangle =J/U\sqrt{\bar n(\bar n+1)}\sum_m \left (\vert m,1\rangle + \vert m,L-1\rangle \right). \nonumber
\end{eqnarray}
To determine the corrections to the particle-hole excitations, one needs to diagonalize $H_{\text{kin}}$ within
the lowest band of excitations. This yields the diagonal basis~\cite{IucciGiamarchi2006,BarmettlerKollath2012}
\begin{align}\label{eqn:Kkbasis}
\vert K, q \rangle=\frac{\sqrt{2}}{L}\sum_{d=1}^{L-1}\sum_{m=1}^Le^{id\theta (K)}\sin(q x_d)e^{iKx_m}\vert m,d\rangle,
\end{align}
where $x_m=am$, $x_d=ad$, $a$ is the lattice spacing and
$\theta(K)=(\bar n+1)\sin(Ka)/[\bar n+(\bar n+1)\cos(Ka)]$. Here $K=2\pi b/(La)$ with $b=1,...,L$ can
be interpreted as a center of mass momentum and $q=\pi l/(aL)$  with $l=1,...,L-1$ is related to the relative
momentum of the excess and hole particles. Note, that a Fourier transform corresponding
to the distance has to be taken for open boundary conditions.
This basis provides the lowest order (zero-order) eigenstates. While the first order
correction to the energy is given by $-2J~r(K)\cos(qa)$ with
$r(K)=\sqrt{(\bar n+1)^2+\bar n^2+2\bar n (\bar n+1)\cos(Ka)}$, such that the energy of
the lowest excitation band becomes
\begin{eqnarray}
  E_{K,q}= U-2J~r(K)\cos(qa) \nonumber
\end{eqnarray}
lifting the degeneracy except for a translational invariance in $K$ by $2\pi/a$. 

\paragraph{Application to the energy absorption rate}
We determine the energy absorption rate within linear response, Eq.~\eqref{eqn:dEdt}, for the excitations
created around the modulation frequency $\hbar\omega \approx U$. To do so, we evaluate
the resonance condition using the energy expressions obtained via
perturbation theory, i.e. $E_0=0$, $E_{K,q}=U-2J~r(K)\cos(qa)$. 
At the considered order, the relevant matrix element is $|\langle \Psi_1|\hat O|\Psi_0\rangle|$ where
$|\Psi_0\rangle=|0\rangle + |\Psi_0^1\rangle +\mathcal O(J^2/U^2)$ and
$|\Psi_1\rangle=|K,q\rangle -(J/U)\sqrt{2\bar n(\bar n+1)}\eta_l\sin(q a)|0\rangle +(J/U)\sum_{\alpha}|\alpha\rangle +\mathcal O(J^2/U^2)$
where $|\alpha \rangle$ are states, in addition to the Fock state $|0\rangle$, that are directly coupled via
the kinetic term to the states $|K,q\rangle$. The squared norm of the transition matrix element simplifies as
$|\langle \Psi_1|\hat O|\Psi_0\rangle|^2=|\langle K,q|\hat O|0\rangle|^2+\mathcal O(J^2/U^2)$, where
$\langle K,q|\hat O|0\rangle=\sqrt{2\bar n (\bar n+1)}\sin(qa)\eta_l \delta_{aK,\pi}$ with $\eta_l=[1-(-1)^l]$.
Using these expressions, the energy absorption rate in the continuum limit, $L\rightarrow \infty$, becomes
\begin{align}\label{eqn:dEdtPT}
\frac{1}{L}\frac{\overline{dE(t)}}{dt}&=\frac{\omega \vert A \vert ^2\bar n(\bar n+1) }{J} \sqrt{1-\left ( \frac{U-\hbar \omega}{2J}\right)^2}.
\end{align}
Thus, absorption occurs in the region $[U-2J,U+ 2J]$ corresponding to the width $4J$ of the lowest
band of excitations for $aK=\pi$. The absorption maximum is located at
$\hbar \omega_{\textrm{peak}}\approx U\left(1+\left (2J/U\right)^2\right)$. 

\subsection{Results in the Mott-insulating phase}\label{sec:MI_results}
\begin{figure}
\includegraphics[width=.99\columnwidth,clip=true]{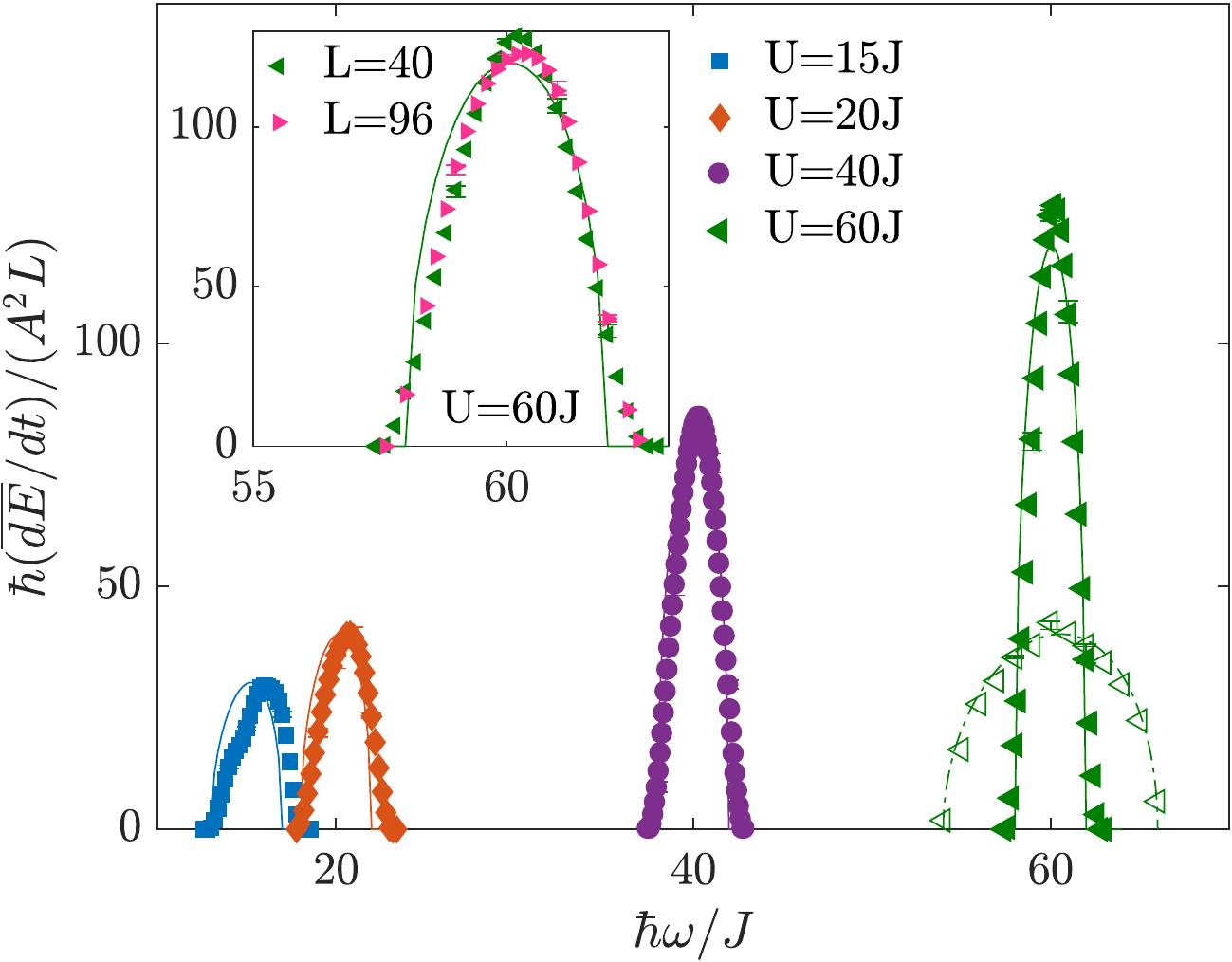}
\caption{Energy absorption rates deep in the Mott insulator for
  a system of size $L=40$ and a modulation amplitude $A=0.01J$. Symbols are t-DMRG results and solid lines show
  the analytical result within perturbation theory (see Eq.~\eqref{eqn:dEdtPT}). For $U=60J$,
  a comparison to the normal lattice modulation is shown (open symbols). The dashed-dotted line is the
  response to normal lattice modulation within perturbation theory~\cite{IucciGiamarchi2006}.
  The inset shows a comparison to a system of size $L=96$ at $U=60J$.
}
\label{fig:MIlargeU}
\end{figure}

\begin{figure}
\includegraphics[width=.99\columnwidth,clip=true]{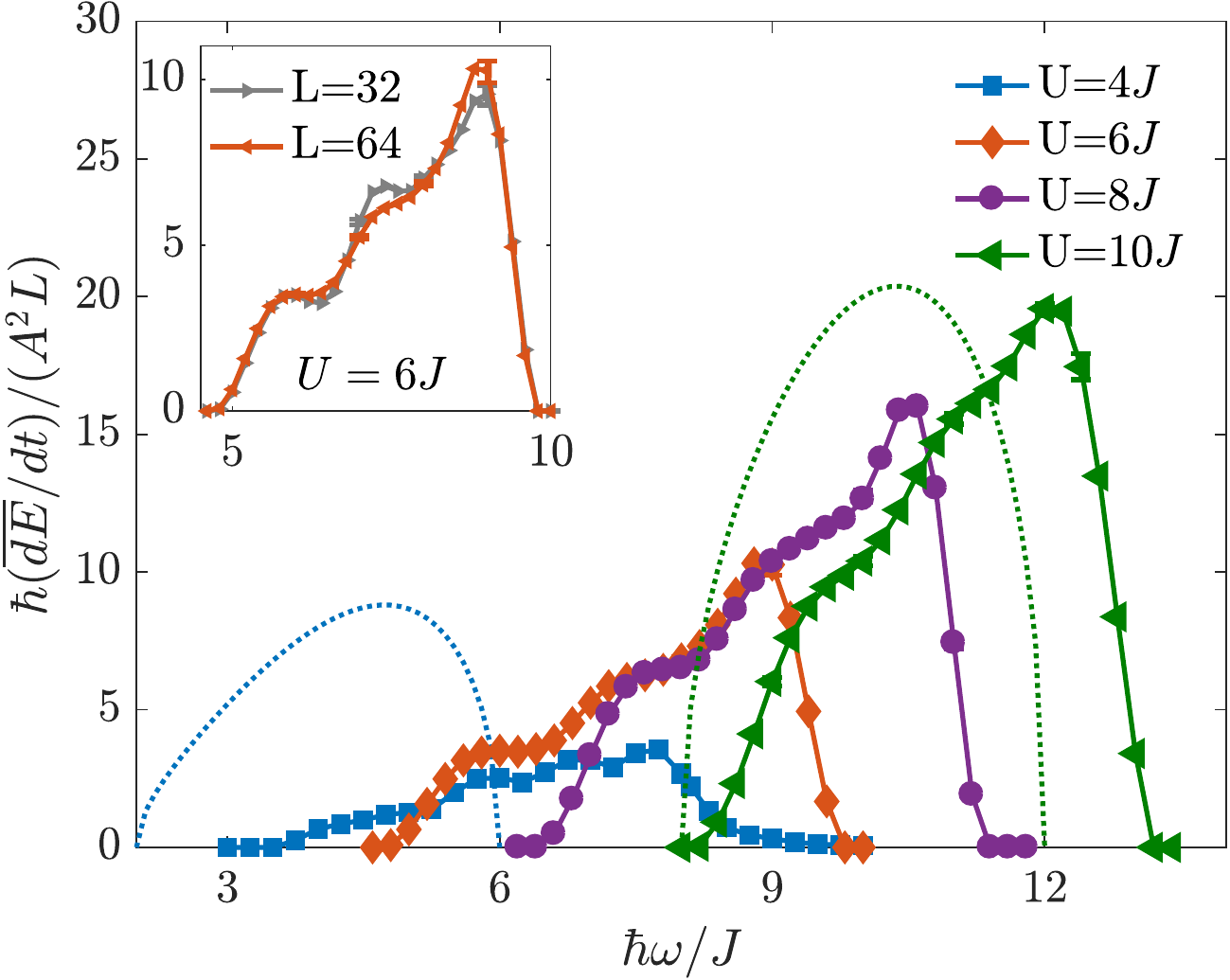}
\caption{Energy absorption rates for intermediate interaction strengths on the Mott-insulating side
  of the phase transition for a system of size $L=64$ and a modulation amplitude $A=0.01J$.
  Symbols are t-DMRG results and solid lines are guide to the eyes. For $U=4J$ and $U=10J$, dotted lines show that the
  analytical predictions of perturbation theory deviate more and more from numerical results as $U$ is approaching the
  phase transition to the superfluid state. The inset shows a comparison to a system size $L=32$ at $U=6J$. Plateaus
  in the absorption rate appear to wash out with increasing system size.
  }
\label{fig:MIsmallU}
\end{figure}

Energy absorption rates obtained from t-DMRG and their comparisons with the
perturbative approach are shown in Figs.~\ref{fig:MIlargeU} and~\ref{fig:MIsmallU} at filling ${\bar n}=1$
for strong and intermediate interactions. For strong interactions,
we find very good agreement between the numerical results obtained within t-DMRG and the
perturbative formula Eq.~\eqref{eqn:dEdtPT}. A sharp and narrow absorption peak is found near the frequency $\sim U/\hbar$.
This peak is almost symmetric at large interaction strength and has a width $\sim4J/\hbar$. It becomes more and more
asymmetric at lower interaction strength. Considering different system sizes (see inset of Fig.~\ref{fig:MIlargeU})
a good convergence is already seen for systems of length $L=40$ and $L=96$.
Only small differences arise near the peak maximum. 

For decreasing interaction strengths, the perturbative approach breaks down as this method
can no longer predict accurately the numerical results.
The peak position obtained from t-DMRG moves to the right of the perturbative prediction and
deviates from the naive expectation of $\hbar \omega_{\textrm{peak}} \approx U$. In fact, for $U \lesssim 15J$,
the peak structure becomes more and more asymmetric with a steepening on the high frequency side.
The support of the peak also appears to change with decreasing interaction strength.
Finally, substructures seem to arise (see inset of Fig.~\ref{fig:MIsmallU}). However,
confidently characterizing these substructures would require
larger system sizes such that we will leave this point for further studies.
Considering decreasing interaction strengths within the Mott insulator approaching the phase transition to the superfluid side,
the peak amplitude drops considerably and its extension to high frequency shrinks. We will comment
further on this behavior in the next section where we study the superfluid response. 

\begin{figure}
\includegraphics[width=.99\columnwidth,clip=true]{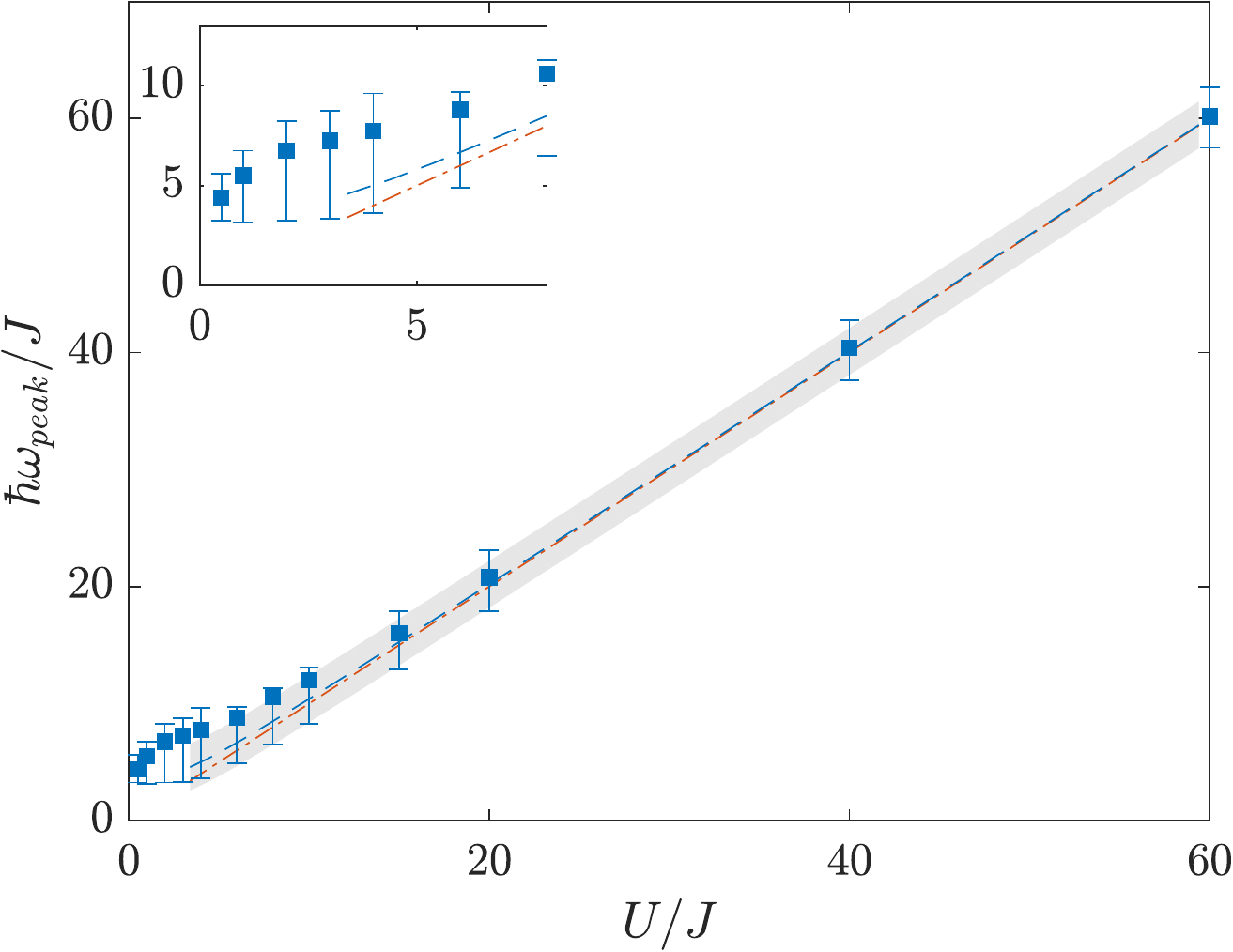}
\caption{The square markers indicate the frequency at which the maximum of the energy absorption rate occurs as a function of interaction
  strength $U$ within t-DMRG using the same parameters as in Figs.~\ref{fig:MIlargeU},~\ref{fig:MIsmallU} and~\ref{fig:SFpeakpos}.
  Error bars indicate the observed bandwidth. We define the bounds as the mean between the frequency for
  which $\overline{dE/dt}/(A^2L)<0.1/\hbar$ and the neighboring frequency for which  $\overline{dE/dt}/(A^2L)>0.1/\hbar$. 
  The dashed blue line indicates the expected frequency within perturbation theory $\hbar\omega_{\text{peak}} \approx U\left(1+(2J/U)^2\right)$ and
  the gray shaded region is the corresponding bandwidth ($=4J$) within perturbation theory. The dash-dotted orange line indicates
  the naive expectation that $\hbar\omega_{\text{peak}} \approx U$. The inset shows a zoom into the small $U$ region.}
\label{fig:MIpeakpos}
\end{figure}

In Fig.~\ref{fig:MIpeakpos}, we plot the frequency at which the maximum energy absorption rate occurs
as a function of the interaction strength. This value calculated using t-DMRG is compared to the perturbative
result and to the naive expectation of $U$. At large interaction strengths, the frequency corresponds to the naive
expectation $\hbar \omega \approx U$ and the width of the energy absorption rate peak is fairly narrow (approx.~$4J$).
Considering smaller interaction strengths $U \approx 10J$, this frequency shifts towards slightly larger values,
but remains close to the value of $U/\hbar$. Finally, for even smaller interaction strengths,
the frequency deviates considerably.
Therefore, the frequency at which the maximum energy absorption rate takes place can be used to infer
the value of the interaction strength in an optical lattice potential down to intermediate interaction strengths. 

This measurement procedure is more accurate than extracting $U$ using normal lattice modulation
(as for example done in Ref.~~\cite{MarkNaegerl2011}) as for the latter the absorption occurs in a larger
region $[U-2J(2\bar n+1),U+2J(2\bar n+1)]$ of minimum width $12J$ at $\bar n=1$ which corresponds
to the lowest band of excitations for $K=0$. The absorption rates Eq.~\eqref{eqn:dEdtPT} at strong interactions $U=60J$
and $\bar n=1$ for both superlattice and normal lattice modulations are shown in Fig.~\ref{fig:MIlargeU}.
The difference in width and amplitude is evident from this comparison.

\section{On the superfluid side of the phase transition}\label{sec:SF}
In this section, we discuss the response of the superfluid to superlattice modulation spectroscopy.
At integer filling the system is superfluid for weak interaction strengths such that the system is
below the phase transition to the Mott insulating state occurring in one-dimension at
$(U/J)_c\approx 3.4$ for $\bar n=1$~\cite{KuehnerMonien2000}.
At incommensurate filling, the system remains superfluid for arbitrary interaction strength.
In suitable limits, we analyze the numerical response and the ones obtained for the Lieb-Liniger
model~\cite{Lieb1963,LiebLiniger1963} and Luttinger liquid \cite{Haldane1981,Giamarchi2003}, both
continuous counterparts to the Bose-Hubbard model. Here, we first summarize the
response expected from these two continuum models, before discussing the numerical results obtained
for the Bose-Hubbard model, and highlighting similarities and differences between the latter and
the continuum models.

\begin{figure}
\includegraphics[width=.99\columnwidth,clip=true]{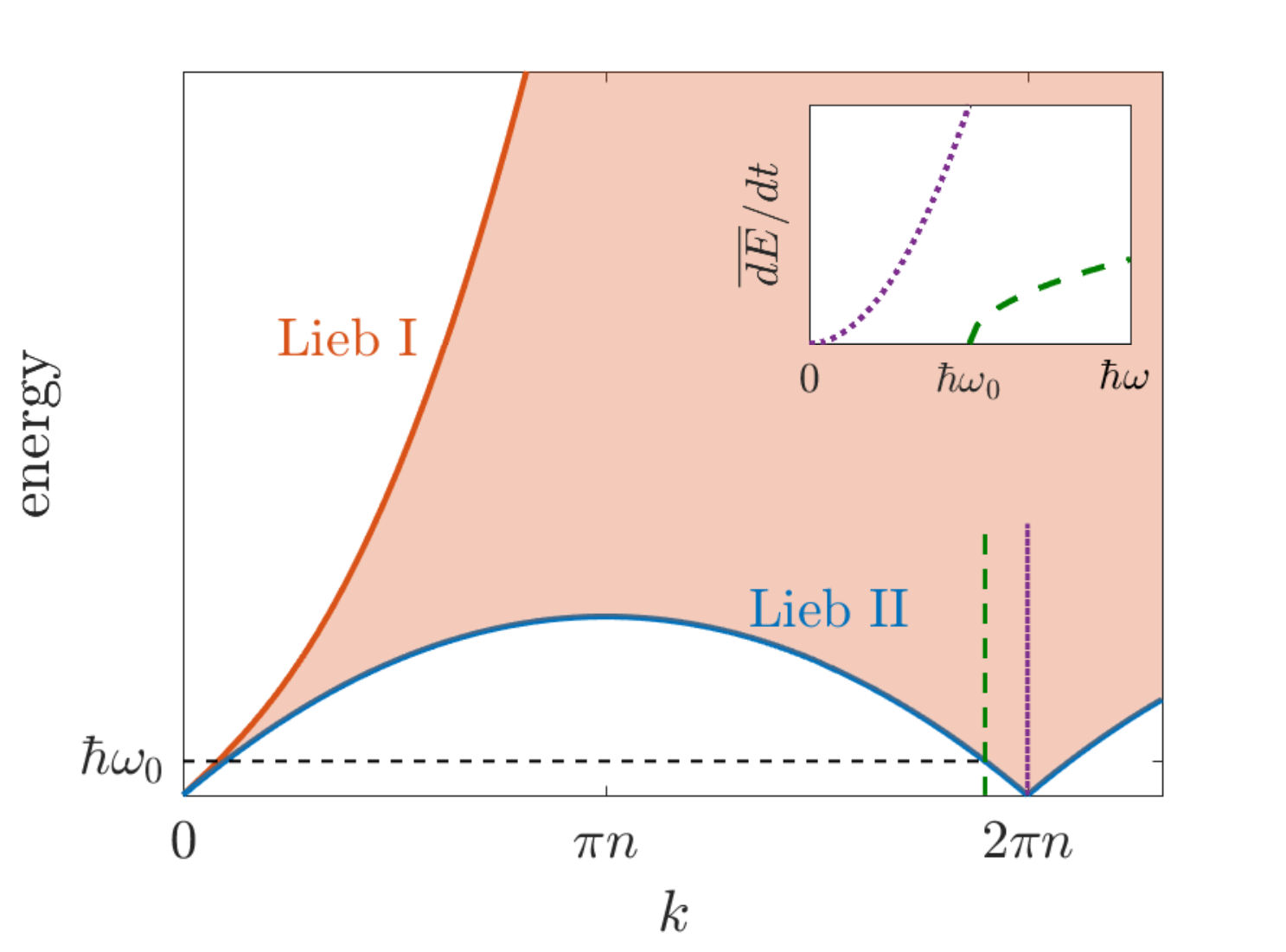}
\caption{Sketch of the excitation spectrum of the Lieb-Liniger model for a given interaction strength $\gamma$.
  The Lieb I mode is sound-like at small momenta and
  becomes particle-like at larger momenta. The Lieb II mode exhibits the same sound-like behavior at small momenta
  but it becomes maximal at $k=\pi n$ and vanishes again at $k=2\pi n$ and reopens at $k>2\pi n$ where $n$ is the
  density. The shaded region represents the continuum of excitations bounded between the two modes. 
  The inset shows a sketch of the onset of the corresponding energy absorption rates within linear response for
  two different densities (using $K=3/2$). The corresponding momentum transfer $\Delta k=\pi/a$ (marked by vertical lines
  in the main plot) either corresponds to a momentum at which the Lieb II mode is finite (dashed green line) which
  leads to a finite onset for the response or to a momentum at which the the energy of the
  Lieb II mode vanishes (dotted purple line) which leads to a finite response at all frequencies.
}
\label{fig:sketch}
\end{figure}

\subsection{Response in the continuum model} \label{sec:SF_LL}

The Lieb-Liniger model is one of the simplest models describing interacting bosonic particles of
mass $M$ in a one-dimensional continuum, assuming a $\delta$-interaction potential of strength $g$,
\begin{align}\label{eqn:dEdtPT}
 \nonumber  H_{LL}=&\int dx \ \Big (\frac{1}{2M}|\partial_x|\Psi(x)|^2
&+ \frac{g}{2}[\Psi^{\dag}(x)]^2[\Psi(x)]^2\Big )  ,
\end{align}
where $\Psi^{(\dag)}(x)$ are the bosonic field operators annihilating (creating) a particle at position
$x$. All quantities are typically expressed in terms of the dimensionless
interaction strength $\gamma=Mg/n$ where $n$ is the density. The Lieb-Liniger model can be obtained from the
Bose-Hubbard model considering its continuum limit by holding $Ja^2$ constant while $a\rightarrow 0$~\cite{KollathZwerger2005_0},
and using the mapping of the parameters $Ja^2=1/2M$, $Ua=g$ and $n=\bar n/a$. The lattice analogue of the dimensionless interaction
is given by $\gamma_{\text{lat}}=(U/J)/2\bar n$. For small values of $\gamma_{\textrm{lat}}$, the Lieb-Liniger model
was found to accurately describe the ground state and some properties of the low energy excitations, such as the sound velocity, of
the Bose-Hubbard model~\cite{KollathZwerger2005_0}. In contrast to the non-integrable Bose-Hubbard model,
the Lieb-Liniger model is Bethe ansatz solvable and therefore many of its properties are well known.
In particular, the model displays two distinct excitations modes, called the Lieb I
and Lieb II modes, sketched in Fig.~\ref{fig:sketch}. The Lieb I mode is sound-like at small momenta and becomes particle-like
at larger momenta. This mode corresponds to the Bogoliubov mode, well known as it arises in the
theory describing weakly interacting Bose gases in higher dimensions. 

A second mode, called Lieb II, arises due to back-scattering in the one-dimensional model.
This mode exhibits the same sound-like behavior at small momenta than the Lieb I as both dispersions have
the same linear slope corresponding to the sound velocity $u$. The Lieb II mode reaches a maximal value
at momentum $k=\pi n$ and vanishes again at $k=2\pi n$. For even larger momenta, a gap reopens in the spectrum.
Such a behavior is typical for one-dimensional models and the low energy excitations around momenta $k=0$ and $k=2\pi n$,
where the dispersion is gapless and linear, are well captured by a bosonization description. 

Within linear response theory, the superlattice modulation operator creates
excitations with a finite momentum transfer $\Delta k= \pi/a$  at a frequency set by the resonance condition
$\hbar \omega=E_\alpha-E_0$, where $E_\alpha$ is the energy of an allowed excitation and $E_0$ the groundstate energy.
Assuming the matrix elements to corresponding momentum transfer to be non-zero, we expect two different kinds of excitations.
The first and generic case occurs at densities where $\Delta k=\pi/a$ corresponds to a momentum value for which
the excitation frequency of the Lieb II mode is finite. Thus, we expect the response in the Lieb-Liniger model
to the superlattice modulation to set in above the corresponding
frequency threshold given by the Lieb II mode, and the upper bound to the frequency is given by the Lieb I mode.
The second type of excitation only occurs if $\Delta k=\pi/a$ is equal to the
momentum $k=2\pi n$ where the energy of the Lieb II mode vanishes. This situation occurs at a density given by $n= 1/(2a)$.
In this case, the superlattice modulation generates excitations even at infinitesimal small frequencies, and the
upper bound is again set by the frequency of the  Lieb I mode. In order to determine the exact
form of the response, the matrix element of the superlattice operator with the particular excitation
need to be computed. Such calculations were performed, for example, in Ref.~\cite{CauxSlavnov2007}
for the single particle spectral function.

These two cases can be further analyzed within a bosonization treatment (see appendix~\ref{sec:appendix} for details of this calculation)
of the low energy excitations. This investigation predicts at the special density point $n= 1/(2a)$ an algebraic onset of
the response for small modulation frequencies $\omega$, i.e.~ 
\begin{equation}
\label{eq:bos_gapless}
 \frac{1}{L}\frac{\overline{dE(t)}}{dt} \propto \omega^{2 K-1}.
\end{equation}
The exponent
is related to the Luttinger liquid exponent $K$. This result implies that the onset becomes slower
with weaker interactions. Additionally, slightly away from this special point where the response is gapless, bosonization
predicts a response above the threshold $\omega_0=u \delta q$  where $ \delta q =\pi/a-2\pi n$
and $u$ is the sound velocity in agreement with the finite frequency of the Lieb II mode.
There the response is given by 
\begin{eqnarray}
  \label{eq:bos_onset}
  \frac{1}{L}\frac{\overline{dE(t)}}{dt} &\propto& \omega A^2\left (\frac{a}{\hbar u}\right )  \\
  &&~\times\left[\left(\frac{\omega a}{2u}\right)^2 - \left(\frac{\delta q a}{2 }\right)^2\right]^{K-1}
  \Theta[\omega^2 -(u\delta q)^2]. \nonumber
\end{eqnarray}
From this expression, one sees that an algebraic onset depending on the Luttinger exponent,
$(K-1)$, is found above the threshold $\omega_0=u \delta q$.
The response predicted by bosonization is exemplified in the inset of Fig.~\ref{fig:sketch} both at the
special gapless point $k=2\pi n$ and slightly away from this point. For other models with long range order,
bosonization predicts distinct features in the response as for example a divergence above a threshold.
One should note that for very low densities bosonization breaks down. 

\subsection{Response of the Bose-Hubbard model in the superfluid phase}\label{sec:SF_results}
\begin{figure}
\includegraphics[width=.99\columnwidth,clip=true]{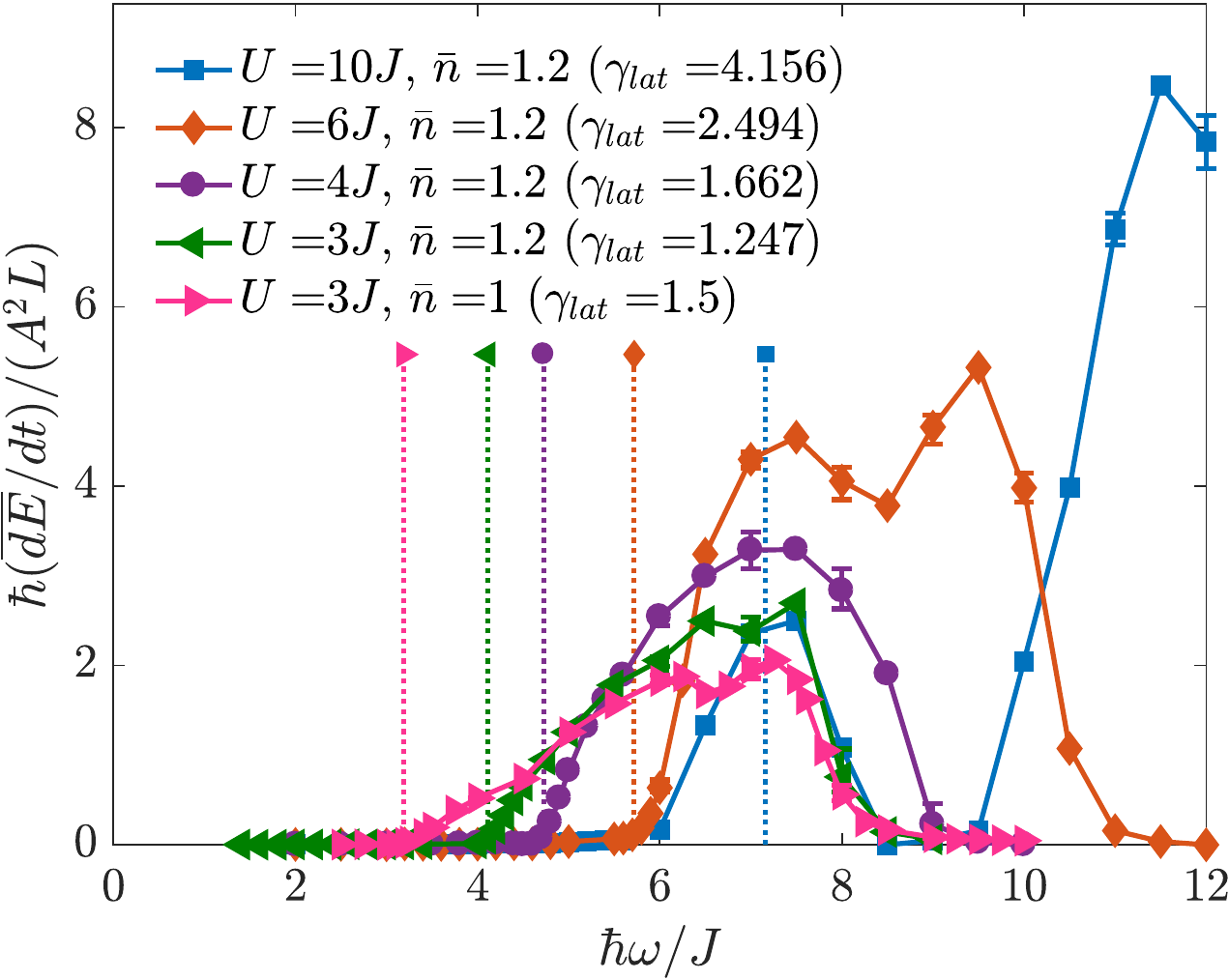}
\caption{The energy absorption rate in the superfluid region for a system of size $L=64$ and a modulation amplitude
  $A=0.01J$ at different interaction strengths. The dotted vertical lines indicate the corresponding energy of the Lieb II
  mode at momentum $ka=\pi$. The fillings and the corresponding continuum densities $n\gtrsim 1/a$  are chosen such that the momentum $ka=\pi$
  appears to the left of the maximum of the Lieb II branch (see Fig.~\ref{fig:sketch}).
  Solid lines are guides to the eye.
}
\label{fig:SFpeakpos}
\end{figure}

We discussed above the expected response of the system to the superlattice modulation in the limit of low energy
using the continuum model. In contrast, we concentrate here on the full Bose-Hubbard model 
for the more generic case of the response occurring above a finite threshold frequency for the densities $n \not= 1/(2a)$ and present the associated
spectral features. The full numerical results for the response of the Bose-Hubbard model are shown
in Fig.~\ref{fig:SFpeakpos} for filling $\bar{n} =1$ and $\bar{n} =1.2$ and for interaction strengths $U$ within the superfluid region. 
For the chosen parameters, the response shows a clear peak structure at finite modulation frequencies.
For low values of $U$ only one peak can be seen in the considered frequency range. At intermediate interaction strength
this peak develops a substructure (see $U=6J$) and then splits up into two separate peaks at larger interaction strength (see $U=10J$).
 
In order to connect these results to the low energy continuum limit, the corresponding values of
$\gamma_\text{lat}$ are given and vertical lines indicate the frequency at which the threshold frequency of the Lieb II mode for $k=\pi/a$ would be located for
the given parameter sets. The onset of the response in the Bose-Hubbard model coincides well with the
predicted Lieb gap at low interaction strength $\gamma_{\textrm{lat}}$. This supports the continuum
description of the low energy excitations of the Bose-Hubbard model.
However, this agreement breaks down for larger values of $\gamma_\text{lat}$ (see $U=10J$)
and when the transferred momentum in units of $k/k_F$ becomes larger.
In the latter case, the difference might solely be due to the slow increase of the typical spectral matrix elements above the
threshold~\cite{CauxSlavnov2007}, such that numerically identifying the location of the onset is difficult. 
At larger interaction strengths additional response features occur. In particular, the observed peak separates into two peaks
one of which lies approximately at $\hbar \omega \approx U$ (see $U=10J$ in Fig.~\ref{fig:SFpeakpos}). We attribute this high energy
peak to particle-hole excitations which arise in the Bose-Hubbard model due to the underlying lattice structure.

\section{Conclusion}\label{sec:con}

In this work, we investigated the response of the one-dimensional Bose-Hubbard model
to superlattice modulation. We demonstrated that
features of the excitation spectrum at finite momenta can be inferred by monitoring the energy
absorption rate during the time-periodic modulation. Using this experimentally realizable setup, we
examined theoretically the response of the system in both the Mott insulating and
superfluid phases.
Deep in the Mott insulator, we found that superlattice modulation creates particle-hole excitations
with finite center of mass momentum $\pi/a$. These excitations are confined to a narrow energy band of width $4J$
well described within a perturbative treatment valid at large interaction strengths. In fact, this
spectral peak is three times narrower than the one observed at zero-momentum transfer.
Superlattice modulation thus enables a more precise experimental calibration of the interaction parameter $U$
than normal lattice modulation would.
In the superfluid phase, the response broadens and different features are displayed. Depending on the filling,
the low energy onset of the response can either be at infinitesimal frequencies
or above a certain threshold which we showed to be related, for low effective interaction strength
$\gamma$, to the spectrum of the Lieb-Liniger model. Moreover, within bosonization, this onset
display an interaction-dependent power-law behavior whose exponent depends on the filling.
For filling ${\bar n} \sim 1$, our numerical results agree well with the onset predicted for the Lieb-Liniger model 
within linear response theory.
Consequently, we demonstrated superlattice modulation spectroscopy to be a versatile and flexible tool to investigate
the finite momentum excitations of strongly correlated quantum phases owing to the momentum transfer introduced
by the dimerization. In fact, this modulation scheme can be extended to an arbitrary momentum transfer $Q$ by modifying the geometry
of the perturbation, i.e.~replacing the dimerization $(-1)^j=\cos(\pi j)$ in Eq.~\eqref{eqn:O} by $\cos(Qaj)$.
This promising extension paves the way to the investigation of more complex lattice models and quantum phases using this
spectroscopic probe.

\appendix
\section{Bosonization approach}\label{sec:appendix} 
In this appendix we sketch the derivation of Eqs.~\eqref{eq:bos_gapless} and \eqref{eq:bos_onset} using a bosonization treatment. The low-energy physics 
of a one-dimensional gas of spinless bosons with repulsive interactions is described by the bosonized Hamiltonian~\cite{Haldane1981_2}
\begin{equation}
  H_0=\int \frac{dx}{2\pi} \left[u K (\pi \Pi(x))^2 + \frac u K (\partial_x \phi(x))^2\right], \nonumber
\end{equation}
where $\phi(x)$ is the bosonic field with conjugate momentum $\pi \Pi(x)$.
The velocity of excitations is given by $u$ and $K$ is the dimensionless Luttinger parameter related to the parameters of the 
original Hamiltonian. In the above formula, and in the remainder of this appendix, we set $\hbar \equiv 1$.
We consider a superlattice modulation with momentum $\pi/a$ 
given by Eq.~\eqref{eqn:O}. In the following we derive a bosonization representation of the corresponding perturbation operator. 
Using the Haldane representation~\cite{Haldane1981_2} of boson annihilation operators,
\begin{equation}
  a_j \sim  e^{i \theta(ja)} \sum_{m=0}^{\infty} A_m \cos 2m (\phi(ja) - \pi n ja), \nonumber
\end{equation}
where $\partial_x \theta (x)= \pi \Pi(x)$, $n$ is the density of atoms, $a$ the lattice spacing, and $A_m$ are amplitudes that depend 
on the details of the microscopic model, we derive
\begin{eqnarray}
 \nonumber  a^\dagger_j a_{j+1} +\text{H.c.}\;~&\sim& C\Pi^2(ja) + D (\partial_x \phi)^2(ja)\\
  \nonumber && + \sum_{m\ne 0}  B_m e^{i2m [\phi(ja) -\pi n a (j + 1/2)]} + \text{H.c.}
\end{eqnarray}
The terms with $C$ and $D$ contribute to the kinetic energy while the terms with $B_m$ contribute to the bond order wave
of wave vector $2\pi m n$. These $B_m$ terms can also be interpreted as the staggered density in the middle of the bond 
$(j,j+1)$.  For $|qa|>1$, the terms proportional to $C$ and $D$ can be neglected and the perturbation operator Eq.~\eqref{eqn:O} becomes
\begin{eqnarray}
  \label{eq:perturbation}
\hat{O} &\sim& \sum_{m\ne 0}  \int dx B'_m e^{i \delta q x} e^{2 i m \phi(x)} \\
&&~~~~~~~~ +  (B'_m)^* e^{-i \delta q x} e^{-2 i m \phi(x)}, \nonumber
\end{eqnarray}
where phases have been absorbed into the phase of $B'_m$  and $ \delta q =\pi/a-2\pi m n$.
The only terms in the sum that may oscillate slowly on the scale of the lattice and contribute at low energies are those with the integer $\bar{m}$ 
being the integer value closest to the value $\frac{1}{2a n}$.
For reasonably large densities $n$, we thus
have \emph{at most} one value of $m=\bar{m}$ for which 
$|\delta q a -2 \pi m n a|\ll 1$ and we obtain the dominant contributions in Eq.~\eqref{eq:perturbation} otherwise the response vanishes.
For the non-vanishing response, the perturbation becomes
\begin{equation}
  H_{\text{pert}} \approx A |B'_{\bar{m}}|  \sin (\omega t)  \int dx \cos (2 \bar{m} \phi(x) - \delta q x + \psi), \nonumber
\end{equation}
where $\psi$ is a phase that can be set to zero by shifting the origin
of coordinates. 
When $A$ is small enough, we can use linear response theory~\cite{LandauLifshitz1959} to calculate the rate of the absorbed energy,
\begin{eqnarray}
  \nonumber  \frac{\overline{dE(t)}}{dt} &\propto& \omega \frac{(A |B'_{\bar m}|)^2}{8} \Big [\mathrm{Im} \chi_{\bar m}(\delta q, \omega + i0_+) \\
&&+ \mathrm{Im} \chi_{\bar m}(-\delta q, \omega + i0_+) \Big], \nonumber
\end{eqnarray}
where $\chi_{\bar m}$ is the retarded response function. To calculate $\chi_{\bar m}$ at zero temperature we use the Matsubara technique.
We have~\cite{Giamarchi2003}
\begin{eqnarray}
  \label{eq:chim}
 \chi_{\bar m}(\delta q, i\omega_n) &=& \int_{-\infty}^{\infty} dx  e^{-i \delta q x} \\
&&\times\int_{-\infty}^{\infty} d\tau e^{i \omega_n \tau} \left(\frac{a^2}{x^2 + (u|\tau|+a)^2}\right)^{\bar m^2 K}. \nonumber
\end{eqnarray}
We first perform the integration over $x$ in Eq.~\eqref{eq:chim} using Eq.~(9.6.25) of Ref.~\cite{AbramowitzStegun1972} and then
we use Eq.~(9.6.23) of Ref.~\cite{AbramowitzStegun1972} to rewrite Eq.~\eqref{eq:chim} and obtain
\begin{eqnarray}
   \nonumber \chi_{\bar m}(\delta q, i\omega_n)&=& \frac{\pi a (a |\delta q|/2)^{2\bar m^2 K -1}} {\Gamma(\bar m^2 K)^2} \\
\nonumber && \times  \int_1^{+\infty} \Bigg \{ dw (w^2
  -1)^{\bar m^2 K-1} e^{-w |\delta q| a}\\
\nonumber && \times  \left(\frac{1}{u|\delta q|w -i
      \omega_n} +  \frac{1}{u|\delta q| w +i \omega_n} \right)\Bigg \}.
\end{eqnarray}
Such an expression allows us to find straightforwardly the analytic
continuation $i\omega_n \to \omega+i0_+$ using
\begin{equation}
  \nonumber \lim_{\epsilon\to 0_+} \frac{1}{x+i\epsilon} = P\left(\frac 1 x
  \right) -\pi \delta(x),
\end{equation}
where $P$ is the principal part, and $\delta$ the Dirac
delta distribution.  We then obtain
\begin{eqnarray} 
  \nonumber \mathrm{Im}\chi_{\bar m}(\delta q,\omega+i0_+) &=&\frac{\pi^2 a^2
    \mathrm{sign}(\omega)}{2 u \Gamma(\bar m^2 K)^2} e^{-\frac{|\omega|a}{u}}\\
\nonumber&& \times  \left[\left(\frac{\omega a}{2 u}\right)^2 - \left(\frac{\delta q
        a}{2 }\right)^2\right]^{\bar m^2 K-1}\\
\nonumber && \times~\Theta[\omega^2 -(u\delta q)^2] ,
\end{eqnarray}
showing that the short distance cutoff in the denominator simply leads
to exponential decay for large $\omega$. For $|\delta q|>0$ at low frequencies, we have
an absorption threshold at $\omega_0=u|\delta q|$. The rate of the absorbed energy
has a divergence at the onset $\omega_0$ when $\bar m^2 K <1$ and a monotonous rise when $\bar m^2 K>1$. In
the case of the Lieb-Liniger gas (or for the Bose-Hubbard model at low
filling), $K>1$ and $\bar m=1$ such that only the rise is seen.


\begin{thebibliography}{46}%
\makeatletter
\providecommand \@ifxundefined [1]{%
 \@ifx{#1\undefined}
}%
\providecommand \@ifnum [1]{%
 \ifnum #1\expandafter \@firstoftwo
 \else \expandafter \@secondoftwo
 \fi
}%
\providecommand \@ifx [1]{%
 \ifx #1\expandafter \@firstoftwo
 \else \expandafter \@secondoftwo
 \fi
}%
\providecommand \natexlab [1]{#1}%
\providecommand \enquote  [1]{``#1''}%
\providecommand \bibnamefont  [1]{#1}%
\providecommand \bibfnamefont [1]{#1}%
\providecommand \citenamefont [1]{#1}%
\providecommand \href@noop [0]{\@secondoftwo}%
\providecommand \href [0]{\begingroup \@sanitize@url \@href}%
\providecommand \@href[1]{\@@startlink{#1}\@@href}%
\providecommand \@@href[1]{\endgroup#1\@@endlink}%
\providecommand \@sanitize@url [0]{\catcode `\\12\catcode `\$12\catcode
  `\&12\catcode `\#12\catcode `\^12\catcode `\_12\catcode `\%12\relax}%
\providecommand \@@startlink[1]{}%
\providecommand \@@endlink[0]{}%
\providecommand \url  [0]{\begingroup\@sanitize@url \@url }%
\providecommand \@url [1]{\endgroup\@href {#1}{\urlprefix }}%
\providecommand \urlprefix  [0]{URL }%
\providecommand \Eprint [0]{\href }%
\providecommand \doibase [0]{http://dx.doi.org/}%
\providecommand \selectlanguage [0]{\@gobble}%
\providecommand \bibinfo  [0]{\@secondoftwo}%
\providecommand \bibfield  [0]{\@secondoftwo}%
\providecommand \translation [1]{[#1]}%
\providecommand \BibitemOpen [0]{}%
\providecommand \bibitemStop [0]{}%
\providecommand \bibitemNoStop [0]{.\EOS\space}%
\providecommand \EOS [0]{\spacefactor3000\relax}%
\providecommand \BibitemShut  [1]{\csname bibitem#1\endcsname}%
\let\auto@bib@innerbib\@empty
\bibitem [{\citenamefont {K\"uhner}\ and\ \citenamefont
  {Monien}(1998)}]{KuehnerMonien1998}%
  \BibitemOpen
  \bibfield  {author} {\bibinfo {author} {\bibfnamefont {T.~D.}\ \bibnamefont
  {K\"uhner}}\ and\ \bibinfo {author} {\bibfnamefont {H.}~\bibnamefont
  {Monien}},\ }\href {\doibase 10.1103/PhysRevB.58.R14741} {\bibfield
  {journal} {\bibinfo  {journal} {Phys. Rev. B}\ }\textbf {\bibinfo {volume}
  {58}},\ \bibinfo {pages} {R14741} (\bibinfo {year} {1998})}\BibitemShut
  {NoStop}%
\bibitem [{\citenamefont {K\"uhner}\ \emph {et~al.}(2000)\citenamefont
  {K\"uhner}, \citenamefont {White},\ and\ \citenamefont
  {Monien}}]{KuehnerMonien2000}%
  \BibitemOpen
  \bibfield  {author} {\bibinfo {author} {\bibfnamefont {T.~D.}\ \bibnamefont
  {K\"uhner}}, \bibinfo {author} {\bibfnamefont {S.~R.}\ \bibnamefont {White}},
  \ and\ \bibinfo {author} {\bibfnamefont {H.}~\bibnamefont {Monien}},\ }\href
  {\doibase 10.1103/PhysRevB.61.12474} {\bibfield  {journal} {\bibinfo
  {journal} {Phys. Rev. B}\ }\textbf {\bibinfo {volume} {61}},\ \bibinfo
  {pages} {12474} (\bibinfo {year} {2000})}\BibitemShut {NoStop}%
\bibitem [{\citenamefont {Zakrzewski}\ and\ \citenamefont
  {Delande}(2008)}]{ZakrzewskiDelande2008}%
  \BibitemOpen
  \bibfield  {author} {\bibinfo {author} {\bibfnamefont {J.}~\bibnamefont
  {Zakrzewski}}\ and\ \bibinfo {author} {\bibfnamefont {D.}~\bibnamefont
  {Delande}},\ }\href {\doibase 10.1063/1.3046265} {\bibfield  {journal}
  {\bibinfo  {journal} {AIP Conference Proceedings}\ }\textbf {\bibinfo
  {volume} {1076}},\ \bibinfo {pages} {292} (\bibinfo {year}
  {2008})}\BibitemShut {NoStop}%
\bibitem [{\citenamefont {Barmettler}\ \emph {et~al.}(2012)\citenamefont
  {Barmettler}, \citenamefont {Poletti}, \citenamefont {Cheneau},\ and\
  \citenamefont {Kollath}}]{BarmettlerKollath2012}%
  \BibitemOpen
  \bibfield  {author} {\bibinfo {author} {\bibfnamefont {P.}~\bibnamefont
  {Barmettler}}, \bibinfo {author} {\bibfnamefont {D.}~\bibnamefont {Poletti}},
  \bibinfo {author} {\bibfnamefont {M.}~\bibnamefont {Cheneau}}, \ and\
  \bibinfo {author} {\bibfnamefont {C.}~\bibnamefont {Kollath}},\ }\href
  {\doibase 10.1103/PhysRevA.85.053625} {\bibfield  {journal} {\bibinfo
  {journal} {Phys. Rev. A}\ }\textbf {\bibinfo {volume} {85}},\ \bibinfo
  {pages} {053625} (\bibinfo {year} {2012})}\BibitemShut {NoStop}%
\bibitem [{\citenamefont {Cazalilla}\ \emph {et~al.}(2011)\citenamefont
  {Cazalilla}, \citenamefont {Citro}, \citenamefont {Giamarchi}, \citenamefont
  {Orignac},\ and\ \citenamefont {Rigol}}]{CazalillaRigol2011}%
  \BibitemOpen
  \bibfield  {author} {\bibinfo {author} {\bibfnamefont {M.~A.}\ \bibnamefont
  {Cazalilla}}, \bibinfo {author} {\bibfnamefont {R.}~\bibnamefont {Citro}},
  \bibinfo {author} {\bibfnamefont {T.}~\bibnamefont {Giamarchi}}, \bibinfo
  {author} {\bibfnamefont {E.}~\bibnamefont {Orignac}}, \ and\ \bibinfo
  {author} {\bibfnamefont {M.}~\bibnamefont {Rigol}},\ }\href {\doibase
  10.1103/RevModPhys.83.1405} {\bibfield  {journal} {\bibinfo  {journal} {Rev.
  Mod. Phys.}\ }\textbf {\bibinfo {volume} {83}},\ \bibinfo {pages} {1405}
  (\bibinfo {year} {2011})}\BibitemShut {NoStop}%
\bibitem [{\citenamefont {L\"auchli}\ and\ \citenamefont
  {Kollath}(2008)}]{LaeuchliKollath2008}%
  \BibitemOpen
  \bibfield  {author} {\bibinfo {author} {\bibfnamefont {A.~M.}\ \bibnamefont
  {L\"auchli}}\ and\ \bibinfo {author} {\bibfnamefont {C.}~\bibnamefont
  {Kollath}},\ }\href {http://stacks.iop.org/1742-5468/2008/i=05/a=P05018}
  {\bibfield  {journal} {\bibinfo  {journal} {Journal of Statistical Mechanics:
  Theory and Experiment}\ }\textbf {\bibinfo {volume} {2008}},\ \bibinfo
  {pages} {P05018} (\bibinfo {year} {2008})}\BibitemShut {NoStop}%
\bibitem [{\citenamefont {Jaksch}\ \emph {et~al.}(1998)\citenamefont {Jaksch},
  \citenamefont {Bruder}, \citenamefont {Cirac}, \citenamefont {Gardiner},\
  and\ \citenamefont {Zoller}}]{JakschZoller1998}%
  \BibitemOpen
  \bibfield  {author} {\bibinfo {author} {\bibfnamefont {D.}~\bibnamefont
  {Jaksch}}, \bibinfo {author} {\bibfnamefont {C.}~\bibnamefont {Bruder}},
  \bibinfo {author} {\bibfnamefont {J.~I.}\ \bibnamefont {Cirac}}, \bibinfo
  {author} {\bibfnamefont {C.~W.}\ \bibnamefont {Gardiner}}, \ and\ \bibinfo
  {author} {\bibfnamefont {P.}~\bibnamefont {Zoller}},\ }\href {\doibase
  10.1103/PhysRevLett.81.3108} {\bibfield  {journal} {\bibinfo  {journal}
  {Phys. Rev. Lett.}\ }\textbf {\bibinfo {volume} {81}},\ \bibinfo {pages}
  {3108} (\bibinfo {year} {1998})}\BibitemShut {NoStop}%
\bibitem [{\citenamefont {Greiner}\ \emph {et~al.}(2002)\citenamefont
  {Greiner}, \citenamefont {Mandel}, \citenamefont {Esslinger}, \citenamefont
  {H{\"a}nsch},\ and\ \citenamefont {Bloch}}]{GreinerBloch2002}%
  \BibitemOpen
  \bibfield  {author} {\bibinfo {author} {\bibfnamefont {M.}~\bibnamefont
  {Greiner}}, \bibinfo {author} {\bibfnamefont {O.}~\bibnamefont {Mandel}},
  \bibinfo {author} {\bibfnamefont {T.}~\bibnamefont {Esslinger}}, \bibinfo
  {author} {\bibfnamefont {T.~W.}\ \bibnamefont {H{\"a}nsch}}, \ and\ \bibinfo
  {author} {\bibfnamefont {I.}~\bibnamefont {Bloch}},\ }\href {\doibase
  10.1038/415039a} {\bibfield  {journal} {\bibinfo  {journal} {Nature}\
  }\textbf {\bibinfo {volume} {415}},\ \bibinfo {pages} {39} (\bibinfo {year}
  {2002})}\BibitemShut {NoStop}%
\bibitem [{\citenamefont {Bloch}\ \emph {et~al.}(2008)\citenamefont {Bloch},
  \citenamefont {Dalibard},\ and\ \citenamefont {Zwerger}}]{BlochZwerger2008}%
  \BibitemOpen
  \bibfield  {author} {\bibinfo {author} {\bibfnamefont {I.}~\bibnamefont
  {Bloch}}, \bibinfo {author} {\bibfnamefont {J.}~\bibnamefont {Dalibard}}, \
  and\ \bibinfo {author} {\bibfnamefont {W.}~\bibnamefont {Zwerger}},\ }\href
  {\doibase 10.1103/RevModPhys.80.885} {\bibfield  {journal} {\bibinfo
  {journal} {Rev. Mod. Phys.}\ }\textbf {\bibinfo {volume} {80}},\ \bibinfo
  {pages} {885} (\bibinfo {year} {2008})}\BibitemShut {NoStop}%
\bibitem [{\citenamefont {Paredes}\ \emph {et~al.}(2004)\citenamefont
  {Paredes}, \citenamefont {Widera}, \citenamefont {Murg}, \citenamefont
  {Mandel}, \citenamefont {F{\"o}lling}, \citenamefont {Cirac}, \citenamefont
  {Shlyapnikov}, \citenamefont {H{\"a}nsch},\ and\ \citenamefont
  {Bloch}}]{ParedesBloch2004}%
  \BibitemOpen
  \bibfield  {author} {\bibinfo {author} {\bibfnamefont {B.}~\bibnamefont
  {Paredes}}, \bibinfo {author} {\bibfnamefont {A.}~\bibnamefont {Widera}},
  \bibinfo {author} {\bibfnamefont {V.}~\bibnamefont {Murg}}, \bibinfo {author}
  {\bibfnamefont {O.}~\bibnamefont {Mandel}}, \bibinfo {author} {\bibfnamefont
  {S.}~\bibnamefont {F{\"o}lling}}, \bibinfo {author} {\bibfnamefont
  {I.}~\bibnamefont {Cirac}}, \bibinfo {author} {\bibfnamefont {G.~V.}\
  \bibnamefont {Shlyapnikov}}, \bibinfo {author} {\bibfnamefont {T.~W.}\
  \bibnamefont {H{\"a}nsch}}, \ and\ \bibinfo {author} {\bibfnamefont
  {I.}~\bibnamefont {Bloch}},\ }\href {http://dx.doi.org/10.1038/nature02530}
  {\bibfield  {journal} {\bibinfo  {journal} {Nature}\ }\textbf {\bibinfo
  {volume} {429}},\ \bibinfo {pages} {277 EP } (\bibinfo {year}
  {2004})}\BibitemShut {NoStop}%
\bibitem [{\citenamefont {St\"oferle}\ \emph {et~al.}(2004)\citenamefont
  {St\"oferle}, \citenamefont {Moritz}, \citenamefont {Schori}, \citenamefont
  {K{\"o}hl},\ and\ \citenamefont {Esslinger}}]{StoeferleEsslinger2004}%
  \BibitemOpen
  \bibfield  {author} {\bibinfo {author} {\bibfnamefont {T.}~\bibnamefont
  {St\"oferle}}, \bibinfo {author} {\bibfnamefont {H.}~\bibnamefont {Moritz}},
  \bibinfo {author} {\bibfnamefont {C.}~\bibnamefont {Schori}}, \bibinfo
  {author} {\bibfnamefont {M.}~\bibnamefont {K{\"o}hl}}, \ and\ \bibinfo
  {author} {\bibfnamefont {T.}~\bibnamefont {Esslinger}},\ }\href {\doibase
  10.1103/PhysRevLett.92.130403} {\bibfield  {journal} {\bibinfo  {journal}
  {Phys. Rev. Lett.}\ }\textbf {\bibinfo {volume} {92}},\ \bibinfo {pages}
  {130403} (\bibinfo {year} {2004})}\BibitemShut {NoStop}%
\bibitem [{\citenamefont {K{\"o}hl}\ \emph {et~al.}(2005)\citenamefont
  {K{\"o}hl}, \citenamefont {Moritz}, \citenamefont {St{\"o}ferle},
  \citenamefont {Schori},\ and\ \citenamefont
  {Esslinger}}]{KoehlEsslinger2005_2}%
  \BibitemOpen
  \bibfield  {author} {\bibinfo {author} {\bibfnamefont {M.}~\bibnamefont
  {K{\"o}hl}}, \bibinfo {author} {\bibfnamefont {H.}~\bibnamefont {Moritz}},
  \bibinfo {author} {\bibfnamefont {T.}~\bibnamefont {St{\"o}ferle}}, \bibinfo
  {author} {\bibfnamefont {C.}~\bibnamefont {Schori}}, \ and\ \bibinfo {author}
  {\bibfnamefont {T.}~\bibnamefont {Esslinger}},\ }\href {\doibase
  10.1007/s10909-005-2273-4} {\bibfield  {journal} {\bibinfo  {journal}
  {Journal of Low Temperature Physics}\ }\textbf {\bibinfo {volume} {138}},\
  \bibinfo {pages} {635} (\bibinfo {year} {2005})}\BibitemShut {NoStop}%
\bibitem [{\citenamefont {Cl\'ement}\ \emph {et~al.}(2009)\citenamefont
  {Cl\'ement}, \citenamefont {Fabbri}, \citenamefont {Fallani}, \citenamefont
  {Fort},\ and\ \citenamefont {Inguscio}}]{ClementInguscio2009}%
  \BibitemOpen
  \bibfield  {author} {\bibinfo {author} {\bibfnamefont {D.}~\bibnamefont
  {Cl\'ement}}, \bibinfo {author} {\bibfnamefont {N.}~\bibnamefont {Fabbri}},
  \bibinfo {author} {\bibfnamefont {L.}~\bibnamefont {Fallani}}, \bibinfo
  {author} {\bibfnamefont {C.}~\bibnamefont {Fort}}, \ and\ \bibinfo {author}
  {\bibfnamefont {M.}~\bibnamefont {Inguscio}},\ }\href {\doibase
  10.1007/s10909-009-0040-7} {\bibfield  {journal} {\bibinfo  {journal}
  {Journal of Low Temperature Physics}\ }\textbf {\bibinfo {volume} {158}},\
  \bibinfo {pages} {5} (\bibinfo {year} {2009})}\BibitemShut {NoStop}%
\bibitem [{\citenamefont {Haller}\ \emph {et~al.}(2010)\citenamefont {Haller},
  \citenamefont {Hart}, \citenamefont {Mark}, \citenamefont {Danzl},
  \citenamefont {Reichs{\"o}llner}, \citenamefont {Gustavsson}, \citenamefont
  {Dalmonte}, \citenamefont {Pupillo},\ and\ \citenamefont
  {N{\"a}gerl}}]{HallerNaegerl2010}%
  \BibitemOpen
  \bibfield  {author} {\bibinfo {author} {\bibfnamefont {E.}~\bibnamefont
  {Haller}}, \bibinfo {author} {\bibfnamefont {R.}~\bibnamefont {Hart}},
  \bibinfo {author} {\bibfnamefont {M.~J.}\ \bibnamefont {Mark}}, \bibinfo
  {author} {\bibfnamefont {J.~G.}\ \bibnamefont {Danzl}}, \bibinfo {author}
  {\bibfnamefont {L.}~\bibnamefont {Reichs{\"o}llner}}, \bibinfo {author}
  {\bibfnamefont {M.}~\bibnamefont {Gustavsson}}, \bibinfo {author}
  {\bibfnamefont {M.}~\bibnamefont {Dalmonte}}, \bibinfo {author}
  {\bibfnamefont {G.}~\bibnamefont {Pupillo}}, \ and\ \bibinfo {author}
  {\bibfnamefont {H.-C.}\ \bibnamefont {N{\"a}gerl}},\ }\href
  {http://dx.doi.org/10.1038/nature09259} {\bibfield  {journal} {\bibinfo
  {journal} {Nature}\ }\textbf {\bibinfo {volume} {466}},\ \bibinfo {pages}
  {597 EP } (\bibinfo {year} {2010})}\BibitemShut {NoStop}%
\bibitem [{\citenamefont {T\"orm\"a}(2015)}]{Toermae2015}%
  \BibitemOpen
  \bibfield  {author} {\bibinfo {author} {\bibfnamefont {P.}~\bibnamefont
  {T\"orm\"a}},\ }in\ \href@noop {} {\emph {\bibinfo {booktitle} {Quantum Gas
  Experiments}}},\ \bibinfo {editor} {edited by\ \bibinfo {editor}
  {\bibfnamefont {P.}~\bibnamefont {T\"orm\"a}}\ and\ \bibinfo {editor}
  {\bibfnamefont {K.}~\bibnamefont {Sengstock}}}\ (\bibinfo  {publisher}
  {Imperial College Press},\ \bibinfo {address} {London},\ \bibinfo {year}
  {2015})\ Chap.~\bibinfo {chapter} {10}, pp.\ \bibinfo {pages}
  {199--250}\BibitemShut {NoStop}%
\bibitem [{\citenamefont {Bissbort}\ \emph {et~al.}(2011)\citenamefont
  {Bissbort}, \citenamefont {G\"otze}, \citenamefont {Li}, \citenamefont
  {Heinze}, \citenamefont {Krauser}, \citenamefont {Weinberg}, \citenamefont
  {Becker}, \citenamefont {Sengstock},\ and\ \citenamefont
  {Hofstetter}}]{BissbortHofstetter2011}%
  \BibitemOpen
  \bibfield  {author} {\bibinfo {author} {\bibfnamefont {U.}~\bibnamefont
  {Bissbort}}, \bibinfo {author} {\bibfnamefont {S.}~\bibnamefont {G\"otze}},
  \bibinfo {author} {\bibfnamefont {Y.}~\bibnamefont {Li}}, \bibinfo {author}
  {\bibfnamefont {J.}~\bibnamefont {Heinze}}, \bibinfo {author} {\bibfnamefont
  {J.~S.}\ \bibnamefont {Krauser}}, \bibinfo {author} {\bibfnamefont
  {M.}~\bibnamefont {Weinberg}}, \bibinfo {author} {\bibfnamefont
  {C.}~\bibnamefont {Becker}}, \bibinfo {author} {\bibfnamefont
  {K.}~\bibnamefont {Sengstock}}, \ and\ \bibinfo {author} {\bibfnamefont
  {W.}~\bibnamefont {Hofstetter}},\ }\href {\doibase
  10.1103/PhysRevLett.106.205303} {\bibfield  {journal} {\bibinfo  {journal}
  {Phys. Rev. Lett.}\ }\textbf {\bibinfo {volume} {106}},\ \bibinfo {pages}
  {205303} (\bibinfo {year} {2011})}\BibitemShut {NoStop}%
\bibitem [{\citenamefont {Fabbri}\ \emph {et~al.}(2015)\citenamefont {Fabbri},
  \citenamefont {Panfil}, \citenamefont {Cl\'ement}, \citenamefont {Fallani},
  \citenamefont {Inguscio}, \citenamefont {Fort},\ and\ \citenamefont
  {Caux}}]{FabbriCaux2015}%
  \BibitemOpen
  \bibfield  {author} {\bibinfo {author} {\bibfnamefont {N.}~\bibnamefont
  {Fabbri}}, \bibinfo {author} {\bibfnamefont {M.}~\bibnamefont {Panfil}},
  \bibinfo {author} {\bibfnamefont {D.}~\bibnamefont {Cl\'ement}}, \bibinfo
  {author} {\bibfnamefont {L.}~\bibnamefont {Fallani}}, \bibinfo {author}
  {\bibfnamefont {M.}~\bibnamefont {Inguscio}}, \bibinfo {author}
  {\bibfnamefont {C.}~\bibnamefont {Fort}}, \ and\ \bibinfo {author}
  {\bibfnamefont {J.-S.}\ \bibnamefont {Caux}},\ }\href {\doibase
  10.1103/PhysRevA.91.043617} {\bibfield  {journal} {\bibinfo  {journal} {Phys.
  Rev. A}\ }\textbf {\bibinfo {volume} {91}},\ \bibinfo {pages} {043617}
  (\bibinfo {year} {2015})}\BibitemShut {NoStop}%
\bibitem [{\citenamefont {Meinert}\ \emph {et~al.}(2015)\citenamefont
  {Meinert}, \citenamefont {Panfil}, \citenamefont {Mark}, \citenamefont
  {Lauber}, \citenamefont {Caux},\ and\ \citenamefont
  {N\"agerl}}]{MeinertNaegerl2015}%
  \BibitemOpen
  \bibfield  {author} {\bibinfo {author} {\bibfnamefont {F.}~\bibnamefont
  {Meinert}}, \bibinfo {author} {\bibfnamefont {M.}~\bibnamefont {Panfil}},
  \bibinfo {author} {\bibfnamefont {M.~J.}\ \bibnamefont {Mark}}, \bibinfo
  {author} {\bibfnamefont {K.}~\bibnamefont {Lauber}}, \bibinfo {author}
  {\bibfnamefont {J.-S.}\ \bibnamefont {Caux}}, \ and\ \bibinfo {author}
  {\bibfnamefont {H.-C.}\ \bibnamefont {N\"agerl}},\ }\href {\doibase
  10.1103/PhysRevLett.115.085301} {\bibfield  {journal} {\bibinfo  {journal}
  {Phys. Rev. Lett.}\ }\textbf {\bibinfo {volume} {115}},\ \bibinfo {pages}
  {085301} (\bibinfo {year} {2015})}\BibitemShut {NoStop}%
\bibitem [{\citenamefont {Menotti}\ \emph {et~al.}(2003)\citenamefont
  {Menotti}, \citenamefont {Kr\"amer}, \citenamefont {Pitaevskii},\ and\
  \citenamefont {Stringari}}]{MenottiStringari2003}%
  \BibitemOpen
  \bibfield  {author} {\bibinfo {author} {\bibfnamefont {C.}~\bibnamefont
  {Menotti}}, \bibinfo {author} {\bibfnamefont {M.}~\bibnamefont {Kr\"amer}},
  \bibinfo {author} {\bibfnamefont {L.}~\bibnamefont {Pitaevskii}}, \ and\
  \bibinfo {author} {\bibfnamefont {S.}~\bibnamefont {Stringari}},\ }\href
  {\doibase 10.1103/PhysRevA.67.053609} {\bibfield  {journal} {\bibinfo
  {journal} {Phys. Rev. A}\ }\textbf {\bibinfo {volume} {67}},\ \bibinfo
  {pages} {053609} (\bibinfo {year} {2003})}\BibitemShut {NoStop}%
\bibitem [{\citenamefont {Batrouni}\ \emph {et~al.}(2005)\citenamefont
  {Batrouni}, \citenamefont {Assaad}, \citenamefont {Scalettar},\ and\
  \citenamefont {Denteneer}}]{BatrouniDenteneer2005}%
  \BibitemOpen
  \bibfield  {author} {\bibinfo {author} {\bibfnamefont {G.~G.}\ \bibnamefont
  {Batrouni}}, \bibinfo {author} {\bibfnamefont {F.~F.}\ \bibnamefont
  {Assaad}}, \bibinfo {author} {\bibfnamefont {R.~T.}\ \bibnamefont
  {Scalettar}}, \ and\ \bibinfo {author} {\bibfnamefont {P.~J.~H.}\
  \bibnamefont {Denteneer}},\ }\href {\doibase 10.1103/PhysRevA.72.031601}
  {\bibfield  {journal} {\bibinfo  {journal} {Phys. Rev. A}\ }\textbf {\bibinfo
  {volume} {72}},\ \bibinfo {pages} {031601} (\bibinfo {year}
  {2005})}\BibitemShut {NoStop}%
\bibitem [{\citenamefont {Reischl}\ \emph {et~al.}(2005)\citenamefont
  {Reischl}, \citenamefont {Schmidt},\ and\ \citenamefont
  {Uhrig}}]{ReischlUhrig2005}%
  \BibitemOpen
  \bibfield  {author} {\bibinfo {author} {\bibfnamefont {A.}~\bibnamefont
  {Reischl}}, \bibinfo {author} {\bibfnamefont {K.~P.}\ \bibnamefont
  {Schmidt}}, \ and\ \bibinfo {author} {\bibfnamefont {G.~S.}\ \bibnamefont
  {Uhrig}},\ }\href {\doibase 10.1103/PhysRevA.72.063609} {\bibfield  {journal}
  {\bibinfo  {journal} {Phys. Rev. A}\ }\textbf {\bibinfo {volume} {72}},\
  \bibinfo {pages} {063609} (\bibinfo {year} {2005})}\BibitemShut {NoStop}%
\bibitem [{\citenamefont {Pupillo}\ \emph {et~al.}(2006)\citenamefont
  {Pupillo}, \citenamefont {Rey},\ and\ \citenamefont
  {Batrouni}}]{PupilloBatrouni2006}%
  \BibitemOpen
  \bibfield  {author} {\bibinfo {author} {\bibfnamefont {G.}~\bibnamefont
  {Pupillo}}, \bibinfo {author} {\bibfnamefont {A.~M.}\ \bibnamefont {Rey}}, \
  and\ \bibinfo {author} {\bibfnamefont {G.~G.}\ \bibnamefont {Batrouni}},\
  }\href {\doibase 10.1103/PhysRevA.74.013601} {\bibfield  {journal} {\bibinfo
  {journal} {Phys. Rev. A}\ }\textbf {\bibinfo {volume} {74}},\ \bibinfo
  {pages} {013601} (\bibinfo {year} {2006})}\BibitemShut {NoStop}%
\bibitem [{\citenamefont {Kr\"amer}\ \emph {et~al.}(2005)\citenamefont
  {Kr\"amer}, \citenamefont {Tozzo},\ and\ \citenamefont
  {Dalfovo}}]{KraemerDalfovo2005}%
  \BibitemOpen
  \bibfield  {author} {\bibinfo {author} {\bibfnamefont {M.}~\bibnamefont
  {Kr\"amer}}, \bibinfo {author} {\bibfnamefont {C.}~\bibnamefont {Tozzo}}, \
  and\ \bibinfo {author} {\bibfnamefont {F.}~\bibnamefont {Dalfovo}},\ }\href
  {\doibase 10.1103/PhysRevA.71.061602} {\bibfield  {journal} {\bibinfo
  {journal} {Phys. Rev. A}\ }\textbf {\bibinfo {volume} {71}},\ \bibinfo
  {pages} {061602} (\bibinfo {year} {2005})}\BibitemShut {NoStop}%
\bibitem [{\citenamefont {Iucci}\ \emph {et~al.}(2006)\citenamefont {Iucci},
  \citenamefont {Cazalilla}, \citenamefont {Ho},\ and\ \citenamefont
  {Giamarchi}}]{IucciGiamarchi2006}%
  \BibitemOpen
  \bibfield  {author} {\bibinfo {author} {\bibfnamefont {A.}~\bibnamefont
  {Iucci}}, \bibinfo {author} {\bibfnamefont {M.~A.}\ \bibnamefont
  {Cazalilla}}, \bibinfo {author} {\bibfnamefont {A.~F.}\ \bibnamefont {Ho}}, \
  and\ \bibinfo {author} {\bibfnamefont {T.}~\bibnamefont {Giamarchi}},\ }\href
  {\doibase 10.1103/PhysRevA.73.041608} {\bibfield  {journal} {\bibinfo
  {journal} {Phys. Rev. A}\ }\textbf {\bibinfo {volume} {73}},\ \bibinfo
  {pages} {041608} (\bibinfo {year} {2006})}\BibitemShut {NoStop}%
\bibitem [{\citenamefont {Kollath}\ \emph {et~al.}(2006)\citenamefont
  {Kollath}, \citenamefont {Iucci}, \citenamefont {Giamarchi}, \citenamefont
  {Hofstetter},\ and\ \citenamefont {Schollw\"ock}}]{KollathSchollwoeck2006}%
  \BibitemOpen
  \bibfield  {author} {\bibinfo {author} {\bibfnamefont {C.}~\bibnamefont
  {Kollath}}, \bibinfo {author} {\bibfnamefont {A.}~\bibnamefont {Iucci}},
  \bibinfo {author} {\bibfnamefont {T.}~\bibnamefont {Giamarchi}}, \bibinfo
  {author} {\bibfnamefont {W.}~\bibnamefont {Hofstetter}}, \ and\ \bibinfo
  {author} {\bibfnamefont {U.}~\bibnamefont {Schollw\"ock}},\ }\href {\doibase
  10.1103/PhysRevLett.97.050402} {\bibfield  {journal} {\bibinfo  {journal}
  {Phys. Rev. Lett.}\ }\textbf {\bibinfo {volume} {97}},\ \bibinfo {pages}
  {050402} (\bibinfo {year} {2006})}\BibitemShut {NoStop}%
\bibitem [{\citenamefont {Sensarma}\ \emph {et~al.}(2011)\citenamefont
  {Sensarma}, \citenamefont {Sengupta},\ and\ \citenamefont
  {Das~Sarma}}]{SensarmaDasSarma2011}%
  \BibitemOpen
  \bibfield  {author} {\bibinfo {author} {\bibfnamefont {R.}~\bibnamefont
  {Sensarma}}, \bibinfo {author} {\bibfnamefont {K.}~\bibnamefont {Sengupta}},
  \ and\ \bibinfo {author} {\bibfnamefont {S.}~\bibnamefont {Das~Sarma}},\
  }\href {\doibase 10.1103/PhysRevB.84.081101} {\bibfield  {journal} {\bibinfo
  {journal} {Phys. Rev. B}\ }\textbf {\bibinfo {volume} {84}},\ \bibinfo
  {pages} {081101} (\bibinfo {year} {2011})}\BibitemShut {NoStop}%
\bibitem [{\citenamefont {Fallani}\ \emph {et~al.}(2007)\citenamefont
  {Fallani}, \citenamefont {Lye}, \citenamefont {Guarrera}, \citenamefont
  {Fort},\ and\ \citenamefont {Inguscio}}]{FallaniInguscio2007}%
  \BibitemOpen
  \bibfield  {author} {\bibinfo {author} {\bibfnamefont {L.}~\bibnamefont
  {Fallani}}, \bibinfo {author} {\bibfnamefont {J.~E.}\ \bibnamefont {Lye}},
  \bibinfo {author} {\bibfnamefont {V.}~\bibnamefont {Guarrera}}, \bibinfo
  {author} {\bibfnamefont {C.}~\bibnamefont {Fort}}, \ and\ \bibinfo {author}
  {\bibfnamefont {M.}~\bibnamefont {Inguscio}},\ }\href {\doibase
  10.1103/PhysRevLett.98.130404} {\bibfield  {journal} {\bibinfo  {journal}
  {Phys. Rev. Lett.}\ }\textbf {\bibinfo {volume} {98}},\ \bibinfo {pages}
  {130404} (\bibinfo {year} {2007})}\BibitemShut {NoStop}%
\bibitem [{\citenamefont {Orso}\ \emph {et~al.}(2009)\citenamefont {Orso},
  \citenamefont {Iucci}, \citenamefont {Cazalilla},\ and\ \citenamefont
  {Giamarchi}}]{OrsoGiamarchi2009}%
  \BibitemOpen
  \bibfield  {author} {\bibinfo {author} {\bibfnamefont {G.}~\bibnamefont
  {Orso}}, \bibinfo {author} {\bibfnamefont {A.}~\bibnamefont {Iucci}},
  \bibinfo {author} {\bibfnamefont {M.~A.}\ \bibnamefont {Cazalilla}}, \ and\
  \bibinfo {author} {\bibfnamefont {T.}~\bibnamefont {Giamarchi}},\ }\href
  {\doibase 10.1103/PhysRevA.80.033625} {\bibfield  {journal} {\bibinfo
  {journal} {Phys. Rev. A}\ }\textbf {\bibinfo {volume} {80}},\ \bibinfo
  {pages} {033625} (\bibinfo {year} {2009})}\BibitemShut {NoStop}%
\bibitem [{\citenamefont {Huber}\ \emph {et~al.}(2007)\citenamefont {Huber},
  \citenamefont {Altman}, \citenamefont {B\"uchler},\ and\ \citenamefont
  {Blatter}}]{HuberBlatter2007}%
  \BibitemOpen
  \bibfield  {author} {\bibinfo {author} {\bibfnamefont {S.~D.}\ \bibnamefont
  {Huber}}, \bibinfo {author} {\bibfnamefont {E.}~\bibnamefont {Altman}},
  \bibinfo {author} {\bibfnamefont {H.~P.}\ \bibnamefont {B\"uchler}}, \ and\
  \bibinfo {author} {\bibfnamefont {G.}~\bibnamefont {Blatter}},\ }\href
  {\doibase 10.1103/PhysRevB.75.085106} {\bibfield  {journal} {\bibinfo
  {journal} {Phys. Rev. B}\ }\textbf {\bibinfo {volume} {75}},\ \bibinfo
  {pages} {085106} (\bibinfo {year} {2007})}\BibitemShut {NoStop}%
\bibitem [{\citenamefont {Endres}\ \emph {et~al.}(2012)\citenamefont {Endres},
  \citenamefont {Fukuhara}, \citenamefont {Pekker}, \citenamefont {Cheneau},
  \citenamefont {Schau{\ss}}, \citenamefont {Gross}, \citenamefont {Demler},
  \citenamefont {Kuhr},\ and\ \citenamefont {Bloch}}]{EndresBloch2012}%
  \BibitemOpen
  \bibfield  {author} {\bibinfo {author} {\bibfnamefont {M.}~\bibnamefont
  {Endres}}, \bibinfo {author} {\bibfnamefont {T.}~\bibnamefont {Fukuhara}},
  \bibinfo {author} {\bibfnamefont {D.}~\bibnamefont {Pekker}}, \bibinfo
  {author} {\bibfnamefont {M.}~\bibnamefont {Cheneau}}, \bibinfo {author}
  {\bibfnamefont {P.}~\bibnamefont {Schau{\ss}}}, \bibinfo {author}
  {\bibfnamefont {C.}~\bibnamefont {Gross}}, \bibinfo {author} {\bibfnamefont
  {E.}~\bibnamefont {Demler}}, \bibinfo {author} {\bibfnamefont
  {S.}~\bibnamefont {Kuhr}}, \ and\ \bibinfo {author} {\bibfnamefont
  {I.}~\bibnamefont {Bloch}},\ }\href {\doibase 10.1038/nature11255} {\bibfield
   {journal} {\bibinfo  {journal} {Nature}\ }\textbf {\bibinfo {volume}
  {487}},\ \bibinfo {pages} {454} (\bibinfo {year} {2012})}\BibitemShut
  {NoStop}%
\bibitem [{\citenamefont {Loida}\ \emph {et~al.}(2015)\citenamefont {Loida},
  \citenamefont {Sheikhan},\ and\ \citenamefont {Kollath}}]{LoidaKollath2015}%
  \BibitemOpen
  \bibfield  {author} {\bibinfo {author} {\bibfnamefont {K.}~\bibnamefont
  {Loida}}, \bibinfo {author} {\bibfnamefont {A.}~\bibnamefont {Sheikhan}}, \
  and\ \bibinfo {author} {\bibfnamefont {C.}~\bibnamefont {Kollath}},\ }\href
  {\doibase 10.1103/PhysRevA.92.043624} {\bibfield  {journal} {\bibinfo
  {journal} {Phys. Rev. A}\ }\textbf {\bibinfo {volume} {92}},\ \bibinfo
  {pages} {043624} (\bibinfo {year} {2015})}\BibitemShut {NoStop}%
\bibitem [{\citenamefont {Loida}\ \emph {et~al.}(2017)\citenamefont {Loida},
  \citenamefont {Bernier}, \citenamefont {Citro}, \citenamefont {Orignac},\
  and\ \citenamefont {Kollath}}]{LoidaKollath2017}%
  \BibitemOpen
  \bibfield  {author} {\bibinfo {author} {\bibfnamefont {K.}~\bibnamefont
  {Loida}}, \bibinfo {author} {\bibfnamefont {J.-S.}\ \bibnamefont {Bernier}},
  \bibinfo {author} {\bibfnamefont {R.}~\bibnamefont {Citro}}, \bibinfo
  {author} {\bibfnamefont {E.}~\bibnamefont {Orignac}}, \ and\ \bibinfo
  {author} {\bibfnamefont {C.}~\bibnamefont {Kollath}},\ }\href {\doibase
  10.1103/PhysRevLett.119.230403} {\bibfield  {journal} {\bibinfo  {journal}
  {Phys. Rev. Lett.}\ }\textbf {\bibinfo {volume} {119}},\ \bibinfo {pages}
  {230403} (\bibinfo {year} {2017})}\BibitemShut {NoStop}%
\bibitem [{\citenamefont {Schollw\"ock}(2011)}]{Schollwoeck2011}%
  \BibitemOpen
  \bibfield  {author} {\bibinfo {author} {\bibfnamefont {U.}~\bibnamefont
  {Schollw\"ock}},\ }\href {\doibase
  http://dx.doi.org/10.1016/j.aop.2010.09.012} {\bibfield  {journal} {\bibinfo
  {journal} {Annals of Physics}\ }\textbf {\bibinfo {volume} {326}},\ \bibinfo
  {pages} {96 } (\bibinfo {year} {2011})}\BibitemShut {NoStop}%
\bibitem [{\citenamefont {Lieb}(1963)}]{Lieb1963}%
  \BibitemOpen
  \bibfield  {author} {\bibinfo {author} {\bibfnamefont {E.~H.}\ \bibnamefont
  {Lieb}},\ }\href {\doibase 10.1103/PhysRev.130.1616} {\bibfield  {journal}
  {\bibinfo  {journal} {Phys. Rev.}\ }\textbf {\bibinfo {volume} {130}},\
  \bibinfo {pages} {1616} (\bibinfo {year} {1963})}\BibitemShut {NoStop}%
\bibitem [{\citenamefont {Lieb}\ and\ \citenamefont
  {Liniger}(1963)}]{LiebLiniger1963}%
  \BibitemOpen
  \bibfield  {author} {\bibinfo {author} {\bibfnamefont {E.~H.}\ \bibnamefont
  {Lieb}}\ and\ \bibinfo {author} {\bibfnamefont {W.}~\bibnamefont {Liniger}},\
  }\href {\doibase 10.1103/PhysRev.130.1605} {\bibfield  {journal} {\bibinfo
  {journal} {Phys. Rev.}\ }\textbf {\bibinfo {volume} {130}},\ \bibinfo {pages}
  {1605} (\bibinfo {year} {1963})}\BibitemShut {NoStop}%
\bibitem [{\citenamefont {Daley}\ \emph {et~al.}(2004)\citenamefont {Daley},
  \citenamefont {Kollath}, \citenamefont {Schollw{\"o}ck},\ and\ \citenamefont
  {Vidal}}]{DaleyVidal2004}%
  \BibitemOpen
  \bibfield  {author} {\bibinfo {author} {\bibfnamefont {A.~J.}\ \bibnamefont
  {Daley}}, \bibinfo {author} {\bibfnamefont {C.}~\bibnamefont {Kollath}},
  \bibinfo {author} {\bibfnamefont {U.}~\bibnamefont {Schollw{\"o}ck}}, \ and\
  \bibinfo {author} {\bibfnamefont {G.}~\bibnamefont {Vidal}},\ }\href
  {http://stacks.iop.org/1742-5468/2004/i=04/a=P04005} {\bibfield  {journal}
  {\bibinfo  {journal} {Journal of Statistical Mechanics: Theory and
  Experiment}\ }\textbf {\bibinfo {volume} {2004}},\ \bibinfo {pages} {P04005}
  (\bibinfo {year} {2004})}\BibitemShut {NoStop}%
\bibitem [{\citenamefont {White}\ and\ \citenamefont
  {Feiguin}(2004)}]{WhiteFeiguin2004}%
  \BibitemOpen
  \bibfield  {author} {\bibinfo {author} {\bibfnamefont {S.~R.}\ \bibnamefont
  {White}}\ and\ \bibinfo {author} {\bibfnamefont {A.~E.}\ \bibnamefont
  {Feiguin}},\ }\href {\doibase 10.1103/PhysRevLett.93.076401} {\bibfield
  {journal} {\bibinfo  {journal} {Phys. Rev. Lett.}\ }\textbf {\bibinfo
  {volume} {93}},\ \bibinfo {pages} {076401} (\bibinfo {year}
  {2004})}\BibitemShut {NoStop}%
\bibitem [{\citenamefont {Sakurai}(1994)}]{Sakurai1994}%
  \BibitemOpen
  \bibfield  {author} {\bibinfo {author} {\bibfnamefont {J.~J.}\ \bibnamefont
  {Sakurai}},\ }\href@noop {} {\emph {\bibinfo {title} {Modern Quantum
  Mechanics}}}\ (\bibinfo  {publisher} {Addison-Wesley Publishing Company},\
  \bibinfo {year} {1994})\BibitemShut {NoStop}%
\bibitem [{\citenamefont {Mark}\ \emph {et~al.}(2011)\citenamefont {Mark},
  \citenamefont {Haller}, \citenamefont {Lauber}, \citenamefont {Danzl},
  \citenamefont {Daley},\ and\ \citenamefont {N{\"a}gerl}}]{MarkNaegerl2011}%
  \BibitemOpen
  \bibfield  {author} {\bibinfo {author} {\bibfnamefont {M.~J.}\ \bibnamefont
  {Mark}}, \bibinfo {author} {\bibfnamefont {E.}~\bibnamefont {Haller}},
  \bibinfo {author} {\bibfnamefont {K.}~\bibnamefont {Lauber}}, \bibinfo
  {author} {\bibfnamefont {J.~G.}\ \bibnamefont {Danzl}}, \bibinfo {author}
  {\bibfnamefont {A.~J.}\ \bibnamefont {Daley}}, \ and\ \bibinfo {author}
  {\bibfnamefont {H.-C.}\ \bibnamefont {N{\"a}gerl}},\ }\href {\doibase
  10.1103/PhysRevLett.107.175301} {\bibfield  {journal} {\bibinfo  {journal}
  {Phys. Rev. Lett.}\ }\textbf {\bibinfo {volume} {107}},\ \bibinfo {pages}
  {175301} (\bibinfo {year} {2011})}\BibitemShut {NoStop}%
\bibitem [{\citenamefont {Haldane}(1981{\natexlab{a}})}]{Haldane1981}%
  \BibitemOpen
  \bibfield  {author} {\bibinfo {author} {\bibfnamefont {F.~D.~M.}\
  \bibnamefont {Haldane}},\ }\href
  {http://stacks.iop.org/0022-3719/14/i=19/a=010} {\bibfield  {journal}
  {\bibinfo  {journal} {Journal of Physics C: Solid State Physics}\ }\textbf
  {\bibinfo {volume} {14}},\ \bibinfo {pages} {2585} (\bibinfo {year}
  {1981}{\natexlab{a}})}\BibitemShut {NoStop}%
\bibitem [{\citenamefont {Giamarchi}(2004)}]{Giamarchi2003}%
  \BibitemOpen
  \bibfield  {author} {\bibinfo {author} {\bibfnamefont {T.}~\bibnamefont
  {Giamarchi}},\ }\href@noop {} {\emph {\bibinfo {title} {Quantum Physics in
  One Dimension}}}\ (\bibinfo  {publisher} {Oxford University Press},\ \bibinfo
  {year} {2004})\BibitemShut {NoStop}%
\bibitem [{\citenamefont {Kollath}\ \emph {et~al.}(2005)\citenamefont
  {Kollath}, \citenamefont {Schollw\"ock}, \citenamefont {von Delft},\ and\
  \citenamefont {Zwerger}}]{KollathZwerger2005_0}%
  \BibitemOpen
  \bibfield  {author} {\bibinfo {author} {\bibfnamefont {C.}~\bibnamefont
  {Kollath}}, \bibinfo {author} {\bibfnamefont {U.}~\bibnamefont
  {Schollw\"ock}}, \bibinfo {author} {\bibfnamefont {J.}~\bibnamefont {von
  Delft}}, \ and\ \bibinfo {author} {\bibfnamefont {W.}~\bibnamefont
  {Zwerger}},\ }\href {\doibase 10.1103/PhysRevA.71.053606} {\bibfield
  {journal} {\bibinfo  {journal} {Phys. Rev. A}\ }\textbf {\bibinfo {volume}
  {71}},\ \bibinfo {pages} {053606} (\bibinfo {year} {2005})}\BibitemShut
  {NoStop}%
\bibitem [{\citenamefont {Caux}\ \emph {et~al.}(2007)\citenamefont {Caux},
  \citenamefont {Calabrese},\ and\ \citenamefont {Slavnov}}]{CauxSlavnov2007}%
  \BibitemOpen
  \bibfield  {author} {\bibinfo {author} {\bibfnamefont {J.-S.}\ \bibnamefont
  {Caux}}, \bibinfo {author} {\bibfnamefont {P.}~\bibnamefont {Calabrese}}, \
  and\ \bibinfo {author} {\bibfnamefont {N.~A.}\ \bibnamefont {Slavnov}},\
  }\href {http://stacks.iop.org/1742-5468/2007/i=01/a=P01008} {\bibfield
  {journal} {\bibinfo  {journal} {Journal of Statistical Mechanics: Theory and
  Experiment}\ }\textbf {\bibinfo {volume} {2007}},\ \bibinfo {pages} {P01008}
  (\bibinfo {year} {2007})}\BibitemShut {NoStop}%
\bibitem [{\citenamefont {Haldane}(1981{\natexlab{b}})}]{Haldane1981_2}%
  \BibitemOpen
  \bibfield  {author} {\bibinfo {author} {\bibfnamefont {F.~D.~M.}\
  \bibnamefont {Haldane}},\ }\href {\doibase 10.1103/PhysRevLett.47.1840}
  {\bibfield  {journal} {\bibinfo  {journal} {Phys. Rev. Lett.}\ }\textbf
  {\bibinfo {volume} {47}},\ \bibinfo {pages} {1840} (\bibinfo {year}
  {1981}{\natexlab{b}})}\BibitemShut {NoStop}%
\bibitem [{\citenamefont {Landau}\ and\ \citenamefont
  {Lifshitz}(1959)}]{LandauLifshitz1959}%
  \BibitemOpen
  \bibfield  {author} {\bibinfo {author} {\bibfnamefont {L.~D.}\ \bibnamefont
  {Landau}}\ and\ \bibinfo {author} {\bibfnamefont {E.~M.}\ \bibnamefont
  {Lifshitz}},\ }\href@noop {} {\emph {\bibinfo {title} {Statistical
  Physics}}}\ (\bibinfo  {publisher} {Pergamon Press (New York)},\ \bibinfo
  {year} {1959})\BibitemShut {NoStop}%
\bibitem [{\citenamefont {Abramowitz}\ and\ \citenamefont
  {Stegun}(1972)}]{AbramowitzStegun1972}%
  \BibitemOpen
  \bibinfo {editor} {\bibfnamefont {M.}~\bibnamefont {Abramowitz}}\ and\
  \bibinfo {editor} {\bibfnamefont {I.}~\bibnamefont {Stegun}},\ eds.,\
  \href@noop {} {\emph {\bibinfo {title} {Handbook of mathematical
  functions}}}\ (\bibinfo  {publisher} {Dover},\ \bibinfo {address} {New
  York},\ \bibinfo {year} {1972})\BibitemShut {NoStop}%
\end{thebibliography}

%

\end{document}